\documentclass[11pt]{article}
\usepackage[utf8]{inputenc}

\usepackage{cite}
\usepackage{amsmath,amscd,amsfonts,mathtools,amssymb}
\usepackage[small,bf,hang]{caption}
\usepackage{slashed}
\usepackage{latexsym,epsfig} 
\usepackage[usenames,dvipsnames]{color}
\usepackage{amssymb}
\usepackage{breqn}
\usepackage{enumerate}
\usepackage{accents}

 \csname
@addtoreset\endcsname{equation}{section}

\newcommand{\ut}[1]{\underaccent{\tilde}{#1}}

\def\moth{\mathsurround=0pt}
\newdimen\zo \zo=0pt

\def\tick{\leaders\hrule height 0.5ex depth 0pt \hskip 0.5pt}
\def\upboxfill{$\moth \setbox\zo\hbox{\tick}%
  \hskip 3pt\hbox to 0pt{$\tick$\hss}\hrulefill \hbox to 7.5pt{$\tick$\hss}$}

\def\dtick{\leaders\hrule height .34pt depth 0.5ex \hskip 0.5pt}
\def\downboxfill{$\moth \setbox\zo\hbox{\dtick}%
  \hskip 2pt\hbox to 0pt{$\dtick$\hss}\hrulefill \hbox to 2pt{$\dtick$\hss}$}

\def\bec{\begin{center}}
\def\ec{\end{center}}

\def\nn{\nonumber}

\def\be{\begin{equation}}
\def\ee{\end{equation}}
\newcommand{\beq}{\begin{equation}\begin{aligned}}
\newcommand{\eeq}{\end{aligned}\end{equation}}
\def\bea{\begin{eqnarray}}
\def\eea{\end{eqnarray}}
\def\ba{\begin{array}}
\def\ea{\end{array}}

\allowdisplaybreaks[2]
\def\hybrid{
        \topmargin -20pt
        \oddsidemargin 0pt
        \headheight 0pt \headsep 0pt
        \textwidth 6.25in 
        \textheight 9.5in 
        \marginparwidth .875in
        \parskip 5pt plus 1pt \jot = 1.5ex}

\hybrid

\begin{document}
\begin{titlepage}

\rightline\today
\begin{center}
\vskip 1.6cm
{\Large \bf {Duality covariant field redefinitions}}\\
 \vskip 2.0cm
{\large {Walter H. Baron}}
\vskip 0.5cm

{\it Instituto de Física La Plata-UNLP-CONICET.\\
and\\
Departamento de Física, Facultad de Ciencias Exactas, Universidad Nacional de La Plata, C.C. 67, (1900) La Plata, Buenos Aires, Argentina.} \\

\vskip 0.5cm

{\small \verb"wbaron@fisica.unlp.edu.ar"}

\vskip 1cm
{\bf Abstract}	
\end{center}

\vskip 0.2cm

\noindent
\begin{narrower}

{\small   We explore the role of the dilaton field on higher-derivative supergravity within the framework of Double Field Theory and use it to fix the Lorentz-noncovariant field redefinitions connecting the metric and dilaton fields with the duality multiplets. }

\end{narrower}

\vskip 1.5cm

\end{titlepage}

\section{Introduction}

 In recent years, there has been a significant progress in the understanding of higher-derivative interactions on supergravity, in particular within the framework of Double Field Theory. That is connected with the fact that the duality symmetries present on theories on toroidal backgrounds constrain the possible interactions even before compactification. In particular, a key result is that T-duality is a symmetry to all orders in the $\alpha'$ expansion of string theory, at tree level \cite{Sen:1991zi}. These symmetries are precisely realized by Double Field Theory (DFT) \cite{DFT} and therefore lead to a very appropriate scheme for discussing higher-derivative corrections. A comprehensive review on that subject can be found in \cite{Lescano:2021lup}.

For instance, the complete Riemann squared terms of the bosonic strings were obtained from duality arguments in \cite{Godazgar}. The first consistent deformation of DFT carrying $\alpha'$ corrections was presented in \cite{HSZ}, and later on it was realized that this theory was a particular point in a two-parameter family of consistent deformations of DFT \cite{Marques:2015vua}. The gauged version of the these theories were obtained in \cite{OddStory} through a Generalized Scherk-Schwarz reduction\cite{GSS}. Alternative approaches that enlarge the duality group while preserving the form of the action where proposed in \cite{OtherApproaches} and interesting applications on cosmological solutions were discussed in \cite{Cosmology}. 
 
 A technique was proposed recently \cite{Baron:2018lve}-\cite{Baron:2020xel} that combines both Lorentz symmetry and T-duality, which presumably allows one to get information on (certain) higher-derivative corrections to the effective actions of string theory in an iterative way. This approach is based on symmetry arguments and not on first principle grounds, but so far all the outcomes are self-consistent and agree with scattering amplitudes, at least up to ${\cal O}(\alpha'{}^2)$. This approach is formulated on DFT, hence a crucial aspect to make contact with first principle computations is to have control on the passage from DFT to the standard SUGRA frame variables.
 
 DFT with strong constraint is basically a T-dual covariant formulation of supergravity. This is written in terms of duality multiplets which admit a simple decomposition in terms of physical fields (dilaton, metric and two-form, if we restrict to the NS-NS sector). Nevertheless as soon as we consider higher-derivative interactions, the simple $GL(D)$ decomposition of the $O(D,D)$ multiplets is no longer identified with the (Lorentz singlets) physical $d.o.f.$ and the connection between SUGRA and DFT frames becomes more and more subtle. The purpose of this article is to shed light on this point. 
    
Inspired by the heterotic supergravity, where anomaly cancellation conditions \cite{Green:1984sg} forced the two-form to get a nontrivial behavior under Lorentz transformations, the authors of \cite{Marques:2015vua} proposed a new duality covariant gauge principle that shed light on $\alpha'$-deformations of DFT through a  {\it Generalized Green Schwarz} ($GGS$) mechanism. 
Actually, it was shown that there are two inequivalent ways to embed the {\it Green Schwarz} ($GS$) transformations in DFT and the requirement of invariance of the action under these transformations led to a two-parameter deformation of the second-derivative action, containing both the heterotic as well as the bosonic string.\footnote{Type II theories are formally included at ${\cal O}(\alpha')$ by setting the parameters to zero, as they get the first nontrivial deformations at 8th derivative order.} 
    
The invariance under $GGS$ and $O(D,D)$ transformations was sufficient to determine the effective action up to quartic order in derivatives. Despite the success of this new gauge principle, it was early realized that the algebra of the proposed $GGS$ only closes at ${\cal O}(\alpha')$ and the absence of a $GS$ like transformation at ${\cal O}(\alpha'{}^2)$ in SUGRA made it difficult to get progress with this method at higher order. 

A proposal to resolve this obstacle was presented in \cite{Baron:2018lve}, where a generalization of the $GGS$ transformations was suggested for the monoparametric case that exactly closes at all orders in derivatives. The prescription from which the perturbative $GGS$ are obtained has its origin on what is was dubbed the {\it Generalized Bergshoeff-deRoo} identification ($GBdRi$). This identification is a dualization of the effective symmetry found by Bergshoeff and de Roo in the heterotic effective action between gauge and (composite) gravity $d.o.f.$, through the  torsionful Lorentz connections \cite{BdR}.
    
Requiring invariance of the action under $GGS$ ensures both Lorentz and duality covariance of the theory\footnote{At this point only those interactions connected by Lorentz and T-duality with the second-derivative action can be accessible, and it is not clear yet whether any other requirement can be exploited to get insight on other interactions, as for instance those being proportional to $\zeta(3)\alpha'{}^3$ present in all string theories. See \cite{Hronek:2021nqk} for a discussion about this point.}. In the frame formulation of DFT T-duality is linearly realized and closes independently order by order in the Lagrangian, Lorentz transformations instead connect interactions with any number of derivatives. Contrarily, in the supergravity framework duality mixes couplings at different orders while Lorentz invariance is realized order by order (with the exception of heterotic case where the Lorentz invariance of the three-form requires higher derivative terms through the Lorentz Chern-Simons). 
    
Even it is formally possible to get the higher derivative interaction terms requiring invariance under $GGS$, in the practice it is not easily implemented. Fortunately, the same technique leading to the $GGS$ can be directly implemented at the level of the action triggering a whole (iterative) tower of interactions, to all orders in powers of $\alpha'$, connected with the second-derivative action through T-duality and Lorentz symmetry. This was successfully implemented in \cite{Baron:2020xel}, not only for the heterotic (monoparametric case) but also for the full family of biparametric duality invariant theories introduced in \cite{Marques:2015vua}. 
    
Besides the symmetry requirements, it is desirable to contrast the results with independent computations, as for instance with scattering amplitudes or $\beta$ function calculations. Interestingly, \cite{Hronek:2021nqk} verified that the $\alpha'{}^2$ interactions obtained in \cite{Baron:2020xel} lead to the expected coefficients for the Riemann cube terms, both for the bosonic and heterotic (zero coefficient in this case).
    
Even though this method allows to get $\alpha'{}^{n}$ deformations within the framework of DFT in an iterative way, the contact with supergravity is not iterative in nature and the necessary Lorentz noncovariant field redefinitions connecting Lorentz and T-duality multiplets must be performed by hand, a work which is hardly implemented beyond ${\cal O}(\alpha'{}^2)$.  
    
In the present article we go back over this point and propose an all orders and iterative field redefinition for the metric and the dilaton fields. The conjectured map relies on scattering amplitude arguments concerning interactions with the dilaton field. We explicitly verified agreement with the expressions in \cite{Baron:2020xel} when reduced to $\alpha'{}^2$. The proposed exact maps are implicitly defined by equations (\ref{gmapBP}), (\ref{phimapBP}), while the explicit expressions up to $\alpha'{}^2$ are displayed on (\ref{gmapalph2}).

The paper is organized as follows. In section 2 we review the Generalized Bergshoeff de Roo identification ($GBdRi$), in section 3 we comment on the possible realization of dilaton free schemes both in SUGRA and DFT. In section 4 we give a proposal for an iterative map connecting physical (Lorentz-invariant) $d.o.f.$ with the duality multiplets. Conclusions are the subject of section 5. 

\section{The Generalized Bergshoeff de Roo identification}
 
In this section we will briefly review the $GBdRi$,  for an extended discussion we recommend the reader to visit the papers \cite{Baron:2018lve} and \cite{Baron:2020xel}. 

T-duality strongly constrains the allowed interactions at low energy and it was proven that there is only a two-parameter family of theories, at fourth-derivative level\cite{Marques:2015vua}, consistent with this symmetry. Particular points of the parameter space are identified with the NS-NS sector of bosonic, heterotic or type II (origin of the parameter space). These deformations are sourced by the two possible ways in which Lorentz transformations can be consistently modified at ${\cal O}(\alpha')$. As opposed to the undeformed Lorentz transformation which is an exact symmetry of the quadratic action, the two-parameter modifications inevitably call for further subleading corrections, as the closure of the new transformations forces the addition of new terms at ${\cal O}(\alpha'{}^2)$. 

Interestingly enough, there is an iterative way to construct a two-parameter minimal completion of the previous deformed transformations, as well as an infinite tower of interaction terms being invariant under such a symmetry. The whole biparametric deformation of DFT is embedded in an artificial extended space, which is infinite dimensional. Regardless of this complication, its dynamics is described by a second-derivative Lagrangian that admits exact local symmetries.

The infinite extra directions are actually related with gauge $d.o.f.$, which are frozen by an exact identification between gauge vectors and spin connections in a duality covariant way. Such a relation, dubbed the {\it Generalized Bergshoeff de Roo identification} ($GBdRi$), relates objects with different number of derivatives triggering all orders $\alpha'$ corrections both in the action and the Lorentz ($GGS$) transformations. 

The gauged DFT we referred to is formally defined on an $2(D+k)$-dimensional extended space (where $k$ is subsequently assumed to be infinite) and the framework on which the $GBdR$ identification is naturally realized is the frame formulation of DFT. The starting point is a gauged flux formulation with rigid ${\cal G}=O(D+k,D+k)=O(D+p+q',D+q+p')$ and local $\underline{\cal H}\times \overline{\cal H}=\underline{O(D+q'-1,p'+1)}\times \overline{O(p+1,D+q-1)}$ symmetry. It is important to keep in mind that the gauge group is not physical but it is artificially introduced to account for higher-derivative deformations of the Lorentz symmetry, in particular the gauge couplings are not related with a compactification scale but with the string length itself through $\alpha'$. 

The index structure that follows from the group decomposition is the following: ${\cal M}\to(M,\hat\mu) = (M,\ut\mu,\tilde\mu)$ denotes the splitting of the rigid group ${\cal G}$ in terms of the fundamental representation of $G=O(D,D)$ and $g=O(k,k)$ respectively, and ${\cal A}=(\underline{\cal A};\overline{\cal A})\to(\underline{a}, \underline{\alpha};\overline{a}, \overline{\alpha})$ parametrizes the decomposition of the local groups $\underline{\cal H}$ into $\underline{H}=\underline{O(D-1,1)}$ and $\underline{h}=\underline{O(q',p')}$ and $\overline{\cal H}$ into $\overline{H}=\overline{O(1,D-1)}$ and $\overline{h}=\overline{O(p,q)}$.

The gauge mechanism introduced is not a minimal gauging procedure as $e.g.$ the heterotic formulation of DFT \cite{FrameDFT}. Here instead the gauging process necessarily introduces scalar fields on top of the gauge vectors which can be embedded on a generalized frame of the $2(D+k)$-dimensional space ${\cal E}_{\cal M}{}^{\cal A}$, admitting the following decomposition
\bea
{\cal E}_M{}^A &=& (\chi^{\frac 1 2 }){}_M{}^N\, E_N{}^A\;, 
\;\;\;\;\;\;\;\;\;\;\;\;\;\;\;\;\;\;
{\cal E}_M{}^{\hat \alpha} = - {\cal A}_M{}^{ {\hat\mu}} \, e_{ {\hat\mu}}{}^{\hat \alpha}\;, \label{FrameExtended}\\
{\cal E}_{ {\hat\mu}}{}^A &=& {\cal A}^M{}_{ {\hat\mu}}\, E_M{}^A\;,
\;\;\;\;\;\;\;\;\;\;\;\;\;\;\;\;\;\;\;\;\;\;\;\;
{\cal E}_{ {\hat\mu}}{}^{\hat \alpha} = (\Box^{\frac 1 2}){}_{ {\hat\mu}}{}^{ {\hat\nu}}\, e_{ {\hat\nu}}{}^{\hat \alpha}\;, \nn
\eea
where the $O(D,D)$-valued $E_{M}{}^{A}$ is identified as the generalized vielbein of the double space and ${\cal A}_{M}{}^{\hat\nu}$ are the (unphysical) gauge vectors. The functions $\chi$ and $\Box$ are defined by
\bea
\chi_{M N} = \eta_{M N} - {\cal A}_M{}^{ {\hat\mu}}\, {\cal A}_{N {\hat\mu}} \ , \ \ \ \ \ \Box_{ {\hat\mu}  {\hat\nu}} = \eta_{ {\hat\mu}  {\hat\nu}} - {\cal A}_{M  {\hat\mu}} \, {\cal A}^M{}_{ {\hat\nu}} \ .
\eea
Here $\eta_{MN}$ is the $O(D,D)$ invariant metric, $\eta_{\hat{\mu} \hat{\nu}}=diag(\eta_{\ut{\mu} \ut{\nu}}, \eta_{\tilde{\mu} \tilde{\nu}})$ with $\eta_{\tilde{\mu}\tilde{\nu}}=\kappa_{\tilde{\mu}\tilde{\nu}}$ and $\eta_{\ut{\mu}\ut{\nu}}=-\kappa_{\ut{\mu}\ut{\nu}}$ where $\kappa$ denotes the killing metric of $\overline{\cal{H}}$ and $\underline{\cal{H}}$, respectively.

Here $e_{ {\hat\mu}}{}^{\hat \alpha}$ is $O(k,k)$-valued, so it can be further parameterized as 
\bea
e_{ {\tilde\mu}}{}^{\overline\alpha} &=& ({\Pi}^{\frac 1 2 })_{ {\tilde\mu}}{}^{ {\tilde\nu}}\, \overline{e}_{ {\tilde\nu}}{}^{\overline\alpha}\;,
\;\;\;\;\;\;\;\;\;\;\;\;\;\;\;\;\;
e_{ {\tilde\mu}}{}^{\underline\alpha} = - \Omega_{ {\tilde\mu}}{}^{{\ut\nu}} \, \underline{e}_{{\ut\nu}}{}^{\underline \alpha}\;, \label{FrameScalar}\\
e_{ {\ut\mu}}{}^{\overline\alpha} &=&  \Omega^{ {\tilde\nu}}{}_{ {\ut\mu}} \, \overline{e}_{ {\tilde\nu}}{}^{\overline \alpha}\;,
\;\;\;\;\;\;\;\;\;\;\;\;\;\;\;\;\;\;\;\;\;\;
e_{ {\ut\mu}}{}^{\underline\alpha} =({\Pi}^{\frac 1 2 })_{ {\ut\mu}}{}^{{\ut\nu}}\, \underline{e}_{{\ut\nu}}{}^{\underline\alpha}\;, \nn
\eea
where $\overline{e}_{ {\tilde\mu}}{}^{\overline\alpha}$ and  $\underline{e}_{ {\ut\mu}}{}^{\underline\alpha}$ are independent $O(p,q)$ and $O(q',p')$ matrices respectively and  
\bea
{\Pi}_{ {\tilde\mu} {\tilde\nu}} = \eta_{ {\tilde\mu} {\tilde\nu}} - \Omega_{ {\tilde\mu}}{}^{{\ut\rho}}\,\Omega_{ {\tilde\nu}}{}_{{\ut\rho}} \ , \;\;\;\;\;\;\;\;\;\;\; {\Pi}_{ {\ut\mu}{\ut\nu}} = \eta_{ {\ut\mu}{\ut\nu}} - \Omega^{ {\tilde\rho}}{}_{ {\ut\mu}}\,\Omega_{ {\tilde\rho} {\ut\nu}} \ .
\eea

Local symmetries of the extended space are parameterized by ${\cal G}$-vectors $\xi^{\cal M}$ that account for both gauge transformations and generalized diffeomorphisms in a duality covariant way and by Lorentz parameters $\Gamma_{\cal A B}$. The generalized frame transforms locally as 
\begin{eqnarray}
\delta {\cal E}_{\cal M}{}^{\cal A}&=&\xi^{\cal N}\partial_{\cal N}{\cal E}_{\cal M}{}^{\cal A} 
+  \left( \partial_{\cal M}\xi^{\cal N} - \partial^{\cal N}\xi_{\cal M} \right) {\cal E}_{\cal N}{}^{\cal A} 
+{\hat f}_{\cal M N}{}^{\cal P} \xi^{\cal N} {\cal E}_{\cal P}{}^{\cal A} 
+ {\cal E}_{\cal M}{}^{\cal B} \Gamma_{\cal B}{}^{\cal A} \;,
\end{eqnarray}
where the information on the gauge group is completely determined by the gaugings ${\hat f}_{\cal M N}{}^{\cal P}$. Concerning the rigid duality group $\cal G$, the generalized vielbein does transform linearly. 

Regarding the generalized dilaton $d$, it turns out to be invariant under Lorentz and the duality group but transforms as a density under generalized diffeomorphisms
\begin{eqnarray}
\delta d &=& \xi^{\cal N} \partial_{N} d - \frac 1 2 \partial_{\cal N} \xi^{\cal N} \ .
\end{eqnarray}

The embedding of DFT with its $\alpha'$-corrections in the $2(D+k)$-dimensional extended space requires fixing extra $d.o.f.$ and symmetries. For instance, setting ${\cal E}_{\ut\mu}{}^{\overline{\alpha}}$ and ${\cal E}_{\tilde\mu}{}^{\underline{\alpha}}$ to zero and  $\underline{e}_{\ut\mu}{}^{\underline\alpha}, \overline{e}_{\tilde\mu}{}^{\overline\alpha}$ to be constant partially break the Lorentz group of the extended space. The remnant symmetry is the usual Lorentz group of the double space, $O(D-1,1)\times O(1,D-1)$. There are still extra $d.o.f.$ of the extended space as compared with those of the NS-NS sector of DFT, and indeed both vectors ${\cal A}^{M}{}_{\hat{\mu}}$ and scalars $\Omega_{\tilde \mu \ut\nu}$ must be frozen. Here is where the $GBdRi$ comes into play: 
\begin{eqnarray}
- g_1\; {\cal E}^{\ut\mu}{}_{\overline{\cal A}} (t_{\ut \mu})_{\underline{\cal B C}}= {\cal F}_{\overline{\cal A}\underline{B C}}\;,
\;\;\;\;\;\;\;\;\;\;\;\;\;\;\;\;\;\;\;\;
- g_2\; {\cal E}^{\tilde\mu}{}_{\underline{\cal A}} (t_{\tilde \mu})_{\overline{\cal B C}}= {\cal F}_{\underline{\cal A}\overline{B C}}\;,
\label{GBdRi}
\end{eqnarray}
where $t_{\ut \mu},\; t_{\tilde \nu}$ are generators of $\underline{\cal H}$ and $\overline{\cal H}$  respectively and $g_1,\;g_2$ their coupling constants. ${\cal F}_{\cal A B C}$ are the generalized fluxes of the extended space
\begin{eqnarray}
{\cal F}_{\cal A B C}= 
3\;\Omega_{[{\cal A B C}]} 
+ {\hat f}_{\cal M N P} {\cal E}^{\cal M}{}_{\cal A} {\cal E}^{\cal N}{}_{\cal B} {\cal E}^{\cal P}{}_{\cal C} \ ,
\;\;\;\;\;\;\;\;\;\;\;\;
{\cal F}_{\cal A}= 2 \;{\cal D}_{\cal A}d -\Omega_{\cal B A}{}^{\cal B}\, , \label{ExtendedFluxes}
\end{eqnarray}
defined in terms of the gaugings ${\hat f}_{\cal M N P}$, generalized Weitzenb\"ock connection $\Omega_{\cal A B C}$ and flat derivatives 
\begin{eqnarray}
\Omega_{\cal A B C}= {\cal D}_{\cal A}{\cal E}^{\cal N}{}_{\cal B} {\cal E}^{\cal P}{}_{\cal C} \; \eta_{\cal N P} \ ,
\;\;\;\;\;\;\;\;\;\;\;\;
{\cal D}_{\cal A}\,= {\cal E}^{\cal M}{}_{\cal A} \partial_{\cal M}\;.
\end{eqnarray}
Equation (\ref{GBdRi}) is supplemented with an identification between gauge and Lorentz parameters of the extended space ($\Gamma_{\underline{\cal B} \overline{\cal C}}\,,\; \Gamma_{\overline{\cal B} \underline{\cal C}}=0$)
\begin{eqnarray}
\Gamma_{\underline{\cal B C}}=-\; g_1 \; \xi^{ {\ut\mu}}\; (t_{ {\ut\mu}})_{\underline{\cal B C}}\,,
\;\;\;\;\;\;\;\;\;\;\;\;\;\;\;\;\;\;\;\;\;
\Gamma_{\overline{\cal B C}}=-\; g_2 \; \xi^{ {\tilde\mu}}\; (t_{ {\tilde\mu}})_{\overline{\cal B C}}\,\;,\label{EABC}
\end{eqnarray}
after which it can be verified that both $l.h.s$ and $r.h.s.$ in (\ref{GBdRi}) operate exactly in the same way under local transformations (here $\widehat {\cal L}_{\xi}$ denotes the generalized Lie derivative).\footnote{It is worth mentioning that this relation truncates the $d.o.f.$ of the extended space but do not impose any condition on the double space. Actually it is after this truncation that the $d.o.f.$ of the extended space exactly matches those of the double space.}
\begin{eqnarray}
\delta {\cal F}_{\underline{\cal A}{\overline{\cal B C}}}&=& \widehat {\cal L}_{\xi}{\cal F}_{\underline{\cal A}{\overline{\cal B C}}}+ {\cal F}_{\underline{\cal D}{\overline{\cal B C}}}\; \Gamma^{\underline{\cal D}}{}_{\underline{\cal A}}-{\cal D}_{\underline {\cal A}}\Gamma_{\overline{\cal B C}} + 2\; {\cal F}_{\underline{\cal A}{\overline{\cal D}[\overline {\cal B}}}\, \Gamma_{\overline{\cal C}]}{}^{\overline{\cal D}}\,,\cr
\delta {\cal F}_{\overline{\cal A}{\underline{\cal B C}}}&=&  \widehat{\cal L}_{\xi}{\cal F}_{\overline{\cal A}{\underline{\cal B C}}} +  {\cal F}_{\overline{\cal D}{\underline{\cal B C}}}\; \Gamma^{\overline{\cal D}}{}_{\overline{\cal A}}-{\cal D}_{\overline {\cal A}}\Gamma_{\underline{\cal B C}} + 2\; {\cal F}_{\overline{\cal A}{\underline{\cal D}[\underline {\cal B}}}\, \Gamma_{\underline{\cal C}]}{}^{\underline{\cal D}}\,,
\end{eqnarray}

Fluxes depend on the derivatives of the generalized frame and so (\ref{GBdRi}) leads to an iterative derivative expansion of the former. If we further take into account the embedding of $E_{M}{}^{A}$ into ${\cal E}_{\cal M}{}^{\cal A}$, then the (exact) generalized diffeomorphism and Lorentz symmetries of extended space trigger a derivative expansion ($\alpha'$-corrections) on the local symmetries for the double space. 

The resulting transformations comprise the usual (uncorrected) generalized diffeomorphism plus a deformed Lorentz symmetry that closes order by order and reduces to the known $GGS$ transformations at ${\cal O}(\alpha')$. Similarly, implementing the $GBdRi$ on the second-derivative action of the extended space leads to an infinite tower of interactions, which again reduces to the known fourth-derivative action at ${\cal O}(\alpha')$,\footnote{It is important to keep in mind that ${\cal L}^{(n)}$ does not necessarily include all possible interactions compatible with scattering amplitudes, it only contains the biparametric deformations of DFT, which is the minimal completion of the second-derivative action that exactly closes under T-duality and the deformed Lorentz transformations ($GGS$) to all orders in derivatives. }
\begin{eqnarray}
{\cal L}\;=\;e^{-2d} \sum_{n=0} {\cal L}^{(n)}\;,\label{fullAction}
\end{eqnarray}
where ${\cal L}^{(n)}\sim \alpha'{}^n$. ${\cal L}^{(0)}= R_1 + R_2$, with
\begin{eqnarray}
R_1&=&2\left(D^{\overline{a}}{F}_{\overline{a}}-
D^{\underline{a}}{F}_{\underline{a}}\right)-\left({F}^{\overline{a}}{F}_{\overline{a}}- {F}^{\underline{a}}{F}_{\underline{a}}\right)\;,\label{R1}\\
R_2&=&\frac12\left({F}_{\underline{a}\overline{b c}} \;{F}^{\underline{a} \overline{b c}} -
{F}_{\overline{a}\underline{b c}}\;{F}^{\overline{a}\underline{b c}} \right)
+\frac16\left( {F}_{\overline{a b c}}\;{F}^{\overline{a b c}}-
 {F}_{\underline{a b c}}\;{F}^{\underline{a b c}}\right).\label{R2}
\end{eqnarray}
Generalized fluxes are defined as
\begin{eqnarray}
F_{ABC}=3\; \Omega_{[ABC]}\, , \;\;\;\;\;\;\;\;\; 
F_{A}=2 D_{A} d - \Omega_{B A}{}^{B}\, , \;\;\;\;\;\;\;\;\; 
\Omega_{ABC}=  \left(D_{A} E^{N}{}_{B}\right) E_{N C}\;,
\end{eqnarray}
and $D_{A}=E^{M}{}_{A} \, \partial_{M}$. 
Alternatively, after integration by parts we can replace $R_1$ with
$\tilde{R}_1$
\begin{eqnarray}
\tilde{R}_1= {F}^{\overline{a}}{F}_{\overline{a}}- {F}^{\underline{a}}{F}_{\underline{a}}\;.\label{tildeR1}
\end{eqnarray}
As we have previously anticipated, the corrections on the second-derivative action are described by two parameters, denoted below by letters $a$ and $b$, whose origin can be traced back to the two coupling constants $g_1$ and $g_2$.
\begin{eqnarray}
{\cal L}^{(1)}&=& a\; {\cal L}^{(1)}_{a} + b\; {\cal L}^{(1)}_{b}
= a \;{\cal L}^{(1)}_{a} +  \left(\begin{matrix} a\leftrightarrow b \cr \{\overline{a}, \overline{b}, \overline{c}, ... \}\leftrightarrow\{\underline{a}, \underline{b}, \underline{c}, ...\}\end{matrix}\right) \;,
\end{eqnarray}
where\footnote{The parameters $a$ and $b$ must not be confused with flat indices $\overline{a},\;\overline{b}\;$ and $\underline{a},\;\underline{b}$. } 
\begin{eqnarray}
{\cal L}^{(1)}_{a}
 &=&2\;D_{\overline{a}}\left(F_{\overline{b}} F^{\overline{a} \underline{c} \underline{d}} F^{\overline{b}}{}_{\underline{c} \underline{d}} \right)\,- \, F_{\overline{a}} F_{\overline{b}} F^{\overline{a} \underline{c} \underline{d}} F^{\overline{b}}{}_{\underline{c} \underline{d}} -D_{\overline{a}}D_{\overline{b}}\left(F^{\overline{a} \underline{c} \underline{d}} F^{\overline{b}}{}_{ \underline{c} \underline{d}} \right)
\cr&&
+ F^{\overline{a} \underline{c} \underline{d}} F^{\overline{b}}{}_{\underline{c} \underline{d}} F_{\overline{a}}{}^{\overline{c} \underline{b}} F_{\overline{b} \overline{c} \underline{b}} - 2 D^{[\overline{a}}F^{\overline{b} ]\underline{c} \underline{d}} D_{\overline{a}}F_{\overline{b} \underline{c} \underline{d}} + \left(F_{\overline{a} \overline{b}\overline{e} } D^{\overline{e}}F^{\overline{a} \underline{c} \underline{d}} - F_{\overline{a} \overline{b} \underline{e}} D^{\underline{e}} F^{\overline{a} \underline{c} \underline{d}}
  \right)  F^{\overline{b}}{}_{\underline{c} \underline{d}} \cr&&
 -4D^{\overline{a}}F^{\overline{b} \underline{a} \underline{c}} F_{\overline{a} \underline{a}}{}^{\underline{b}} F_{\overline{b} \underline{c} \underline{b}} 
+  4/3 F_{\overline{a}}{}^{\underline{a} \underline{c}} F_{\overline{c} \underline{a}}{}^{\underline{b}} F_{\overline{b} \underline{c} \underline{b}} F^{\overline{a} \overline{c} \overline{b}} 
+2 F_{\overline{a}}{}^{\underline{a} \underline{c}} F_{\overline{b}}{}_{\underline{c}}{}^{\underline{b}} F^{\overline{a} }{}_{\underline{d}[\underline{a}} F^{\overline{b}\underline{d} }{}_{\underline{b}]}\;.\label{L1}
\end{eqnarray}

Similarly 
\begin{eqnarray}
 {\cal L}^{(2)}= a^2 {\cal L}^{(0,2)} + a b {\cal L}^{(1,1)} -   \left(\begin{matrix} a\leftrightarrow b \cr \{\overline{a}, \overline{b}, \overline{c}, ... \}\leftrightarrow\{\underline{a}, \underline{b}, \underline{c}, ...\}\end{matrix}\right) \;. 
\end{eqnarray}

The interactions can be further organized according to the presence or absence of dilaton fluxes, $F_{A}$: ${\cal L}^{(r,s)}={\cal L}_{\Phi}^{(r,s)}+{\cal L}_{\not\Phi}^{(r,s)}$. The complete expression is explicitly displayed in \cite{Baron:2020xel}. Here we only exhibit the ${\cal L}_{\Phi}^{(r,s)}$ piece which will be used later on
\begin{eqnarray}
 {\cal L}_{\Phi}^{(0,2)}&=&
 F_{\overline{d}} \left(\vphantom{e^{frac12}}\right.
-2D^{\underline{b}}F^{\overline{d} \underline{e} \underline{f}} D^{\overline{c}}D_{\underline{b}}F_{\overline{c} \underline{e} \underline{f}} 
-2D^{\underline{b}}F^{\overline{a} \underline{e} \underline{f}} D_{\overline{a}}D_{\underline{b}}F^{\overline{d}}{}_{\underline{e} \underline{f}} 
-4D^{\overline{c}}D^{\overline{e}}F_{\overline{c}}{}^{\underline{c} \underline{e}}  F^{\overline{d}}{}_{\underline{c} \underline{f}} F_{\overline{e} \underline{e}}{}^{\underline{f}}\cr
&&
+\, 2D^{\overline{b}}D_{\overline{b}}F^{\overline{e} \underline{c} \underline{e}}  F^{\overline{d}}{}_{\underline{c}}{}^{\underline{f}} F_{\overline{e} \underline{e} \underline{f}} + 2D^{\overline{b}}D^{\underline{f}}F_{\overline{b}}{}^{\underline{d} \underline{e}} F^{\overline{e}}{}_{\underline{d} \underline{e}} F^{\overline{d}}{}_{\overline{e} \underline{f}}
+ 2D^{\underline{f}}F^{\overline{a} \underline{d} \underline{e}} D_{\overline{a}}F^{\overline{e}}{}_{\underline{d} \underline{e}} F^{\overline{d}}{}_{\overline{e} \underline{f}}\cr
&& 
-4D^{\overline{e}}D^{\overline{f}}F^{\overline{d} \underline{c} \underline{e}} F_{\overline{e} \underline{c}}{}^{\underline{f}} F_{\overline{f} \underline{e} \underline{f}} 
+ 2D^{\overline{e}}D^{\overline{d}}F^{\overline{f} \underline{c} \underline{e}} F_{\overline{e} \underline{c}}{}^{\underline{f}} F_{\overline{f} \underline{e} \underline{f}} 
+ 2D^{\overline{e}}D^{\underline{f}}F^{\overline{d} \underline{d} \underline{e}} F^{\overline{f}}{}_{\underline{d} \underline{e}} F_{\overline{e} \overline{f} \underline{f}}\cr
&& 
 + 4D^{\overline{c}}F_{\overline{c}}{}^{\underline{c} \underline{e}} D^{\overline{f}}F^{\overline{d}}{}_{\underline{c} \underline{f}} F_{\overline{f} \underline{e}}{}^{\underline{f}}
 -2D^{\overline{c}}F_{\overline{c}}{}^{\underline{c} \underline{e}} D^{\overline{d}}F^{\overline{f}}{}_{\underline{c}}{}^{\underline{f}} F_{\overline{f} \underline{e} \underline{f}} 
 + 8D^{\overline{d}}F^{\overline{d} \underline{c} \underline{e}} D^{\overline{f}}F_{\overline{d} \underline{c}}{}^{\underline{f}} F_{\overline{f} \underline{e} \underline{f}}\cr
 && 
 -2D^{\overline{d}}F^{\overline{e} \underline{c} \underline{e}} D^{\overline{f}}F_{\overline{e} \underline{c}}{}^{\underline{f}} F_{\overline{f} \underline{e} \underline{f}}
 -2D^{\overline{b}}F^{\overline{d} \underline{c} \underline{e}} D_{\overline{b}}F^{\overline{f}}{}_{\underline{c}}{}^{\underline{f}} F_{\overline{f} \underline{e} \underline{f}}
 + 2D^{\underline{d}}F^{\overline{d} \underline{e} \underline{f}} D^{\overline{c}}F_{\overline{c}}{}^{\overline{f}}{}_{\underline{d}} F_{\overline{f} \underline{e} \underline{f}}\cr
 && 
 + 2D^{\underline{d}}F^{\overline{a} \underline{e} \underline{f}} D_{\overline{a}}F^{\overline{d} \overline{f}}{}_{\underline{d}} F_{\overline{f} \underline{e} \underline{f}} 
 + 2D^{\underline{f}}F^{\overline{d} \underline{d} \underline{e}} D^{\overline{e}}F^{\overline{f}}{}_{\underline{d} \underline{e}} F_{\overline{e} \overline{f} \underline{f}}
 -2D^{\overline{b}}F_{\overline{b}}{}^{\overline{f} \overline{g}} F^{\overline{d} \underline{c} \underline{e}} F_{\overline{f} \underline{c}}{}^{\underline{f}} F_{\overline{g} \underline{e} \underline{f}}\cr
 && 
 -2D^{\overline{b}}F_{\overline{b}}{}^{\overline{e} \underline{f}} F_{\overline{e}}{}^{\underline{d} \underline{e}} F^{\overline{g}}{}_{\underline{d} \underline{e}} F^{\overline{d}}{}_{ \overline{g} \underline{f}}
 + 4D^{\overline{e}}F^{\overline{f} \underline{c} \underline{e}} F^{\overline{d}}{}_{\underline{c}}{}^{\underline{f}} F^{\overline{g}}{}_{\underline{e} \underline{f}} F_{\overline{e} \overline{f} \overline{g}}
 -8D^{\overline{c}}F^{\overline{e} \underline{c} \underline{e}} F^{\overline{d}}{}_{\underline{c}}{}^{\underline{g}} F_{\overline{c} \underline{e}}{}^{\underline{h}} F_{\overline{e} \underline{g} \underline{h}} 
 \cr
 && 
 + 4D^{\overline{e}}F^{\overline{c} \underline{c} \underline{e}} F^{\overline{d}}{}_{\underline{c}}{}^{\underline{g}} F_{\overline{c} \underline{e}}{}^{\underline{h}} F_{\overline{e} \underline{g} \underline{h}}
 + 4D^{\overline{b}}F_{\overline{b}}{}^{\underline{c} \underline{e}} F^{\overline{d}}{}_{\underline{c}}{}^{\underline{g}} F^{\overline{e}}{}_{\underline{e}}{}^{\underline{h}} F_{\overline{e} \underline{g} \underline{h}} 
 + D^{\overline{c}}F^{\overline{e} \underline{c} \underline{d}} F^{\overline{d}}{}_{\underline{c} \underline{d}} F_{\overline{c}}{}^{\underline{g} \underline{h}} F_{\overline{e} \underline{g} \underline{h}} \cr
 && 
 + 4D^{\overline{c}}F^{\overline{e} \underline{e} \underline{g}} F^{\overline{d}\underline{f} \underline{h}} F_{\overline{c} \underline{e} \underline{f}} F_{\overline{e} \underline{g} \underline{h}}
 -4D^{\overline{b}}F_{\overline{b}}{}^{\underline{e} \underline{g}} F^{\overline{d} \underline{f} \underline{h}} F^{\overline{e}}{}_{\underline{e} \underline{f}} F_{\overline{e} \underline{g} \underline{h}} 
+ D^{\overline{c}}F^{\overline{e} \underline{e} \underline{f}} F^{\overline{d} \underline{g} \underline{h}} F_{\overline{c} \underline{e} \underline{f}} F_{\overline{e} \underline{g} \underline{h}}\cr
 && 
 + D^{\overline{b}}F_{\overline{b}}{}^{\underline{e} \underline{f}} F^{\overline{d} \underline{g} \underline{h}} F^{\overline{e}}{}_{\underline{e} \underline{f}} F_{\overline{e} \underline{g} \underline{h}}
 -2D^{\overline{f}}F^{\overline{g} \underline{c} \underline{d}} F^{\overline{e}}{}_{\underline{c} \underline{d}} F^{\overline{d}}{}_{\overline{e}}{}^{\underline{f}} F_{\overline{f} \overline{g} \underline{f}}
-2D^{\overline{f}}F^{\overline{e} \underline{c} \underline{d}} F^{\overline{g}}{}_{\underline{c} \underline{d}} F^{\overline{d}}{}_{\overline{e}}{}^{\underline{f}} F_{\overline{f} \overline{g} \underline{f}}\cr
 && 
 + 4D^{\overline{c}}F^{\overline{f} \underline{c} \underline{e}} F_{\overline{c} \underline{c}}{}^{\underline{f}} F^{\overline{g}}{}_{\underline{e} \underline{f}} F^{\overline{d}}{}_{\overline{f} \overline{g}}
 -2D^{\overline{b}}F_{\overline{b}}{}^{\underline{c} \underline{e}} F^{\overline{f}}{}_{\underline{c}}{}^{\underline{f}} F^{\overline{g}}{}_{\underline{e} \underline{f}} F^{\overline{d}}{}_{\overline{f} \overline{g}}
 -2D^{\overline{f}}F^{\overline{d} \overline{g} \overline{h}}  F_{\overline{f}}{}^{\underline{c} \underline{e}} F_{\overline{g} \underline{c}}{}^{\underline{f}} F_{\overline{h} \underline{e} \underline{f}} \cr
&& 
-2D^{\overline{g}}F^{\overline{d} \overline{e} \underline{f}} F_{\overline{e}}{}^{\underline{d} \underline{e}} F^{\overline{h}}{}_{\underline{d} \underline{e}} F_{\overline{g} \overline{h} \underline{f}}
-2D^{\overline{f}}F^{\overline{d} \underline{c} \underline{e}} F^{\overline{g}}{}_{\underline{c}}{}^{\underline{f}} F^{\overline{h}}{}_{\underline{e} \underline{f}} F_{\overline{f} \overline{g} \overline{h}} 
+ 4D^{\overline{c}}F^{\overline{d} \underline{c} \underline{e}} F_{\overline{c} \underline{c}}{}^{\underline{g}} F^{\overline{f}}{}_{\underline{e}}{}^{\underline{h}} F_{\overline{f} \underline{g} \underline{h}}\cr
&& 
-4D^{\overline{f}}F^{\overline{d} \underline{c} \underline{e}} F^{\overline{e}}{}_{\underline{c}}{}^{\underline{g}} F_{\overline{e} \underline{e}}{}^{\underline{h}} F_{\overline{f} \underline{g} \underline{h}} 
+ D^{\overline{e}}F^{\overline{d} \underline{c} \underline{d}} F^{\overline{f}}{}_{\underline{c} \underline{d}} F_{\overline{e}}{}^{\underline{g} \underline{h}} F_{\overline{f} \underline{g} \underline{h}} \left. \vphantom{2^{\frac12}}\right)
\cr
&+&
D_{\overline{a}}F_{\overline{b}} \left(\vphantom{2^{\frac12}}\right. -2  D^{\underline{c}}F^{\overline{a} \underline{e} \underline{f}} D_{\underline{c}}F^{\overline{b}}{}_{\underline{e} \underline{f}} 
-4 D^{\overline{e}}F^{\overline{b} \underline{c} \underline{e}} F^{\overline{a}}{}_{\underline{c}\underline{f}} F_{\overline{e} \underline{e}}{}^{\underline{f}}
-4 D^{\overline{e}}F^{\overline{a} \underline{c} \underline{e}} F^{\overline{b}}{}_{\underline{c} \underline{f}} F_{\overline{e} \underline{e}}{}^{\underline{f}}  \cr 
&&
+ 2 D^{\overline{a}}F^{\overline{e} \underline{c} \underline{e}} F^{\overline{b}}{}_{\underline{c} \underline{f}} F_{\overline{e} \underline{e}}{}^{\underline{f}} 
+ 2 D^{\overline{b}}F^{\overline{e} \underline{c} \underline{e}} F^{\overline{a}}{}_{\underline{c} \underline{f}} F_{\overline{e} \underline{e}}{}^{\underline{f}} 
+ 2 D^{\underline{f}}F^{\overline{a} \underline{d} \underline{e}} F^{\overline{e}}{}_{\underline{d} \underline{e}} F^{\overline{b}}{}_{\overline{e} \underline{f}}
\cr
&&
+ 2 D^{\underline{f}}F^{\overline{b} \underline{d} \underline{e}} F^{\overline{f}}{}_{\underline{d} \underline{e}} F^{\overline{a}}{}_{\overline{f} \underline{f}} 
 -2 F^{\overline{a} \underline{c} \underline{e}} F^{\overline{f}}{}_{\underline{c}}{}^{\underline{f}} F^{\overline{g}}{}_{\underline{e} \underline{f}} F^{\overline{b}}{}_{\overline{f} \overline{g}} 
-2 F^{\overline{e} \underline{c} \underline{d}} F^{\overline{g}}{}_{\underline{c} \underline{d}} F^{\overline{a} \overline{g} \underline{f}} F^{\overline{b} \overline{e}}{}^{\underline{f}} \cr
&&
 + 4 F^{\overline{a} \underline{c}}{}^{\underline{g}} F^{\overline{e}}{}_{\underline{e}}{}^{\underline{h}} F_{\overline{e} \underline{g} \underline{h}} F^{\overline{b} \underline{c} \underline{e}}
 -4 F^{\overline{d}}{}_{\underline{c}}{}^{\underline{g}} F_{\overline{d} \underline{e}}{}^{\underline{h}} F^{\overline{a}}{}_{\underline{g} \underline{h}} F^{\overline{b} \underline{c} \underline{e}} 
+  F^{\overline{e}}{}_{\underline{c} \underline{d}} F^{\overline{a} \underline{g} \underline{h}} F_{\overline{e} \underline{g} \underline{h}} F^{\overline{b}\underline{c} \underline{d}}\cr
&&
-2 F^{\overline{f}}{}_{\underline{c}}{}^{\underline{f}} F^{\overline{g}}{}_{\underline{e} \underline{f}} F^{\overline{b} \underline{c} \underline{e}} F^{\overline{a}}{}_{\overline{f} \overline{g}} \left.\vphantom{2^{\frac12}}\right)
 \cr
 &&
+ F_{\overline{a}} F_{\overline{b}} \left(\vphantom{2^{\frac12}}\right. D^{\underline{c}}F^{\overline{a} \underline{e} \underline{f}} D_{\underline{c}}F^{\overline{b}}{}_{\underline{e} \underline{f}} 
  -2 D^{\underline{f}}F^{\overline{a} \underline{d} \underline{e}} F^{\overline{e}}{}_{\underline{d} \underline{e}} F^{\overline{b}}{}_{\overline{e} \underline{f}} 
 -2 D^{\overline{a}}F^{\overline{e} \underline{c} \underline{e}} F^{\overline{b}}{}_{\underline{c} \underline{f}} F_{\overline{e} \underline{e}}{}^{\underline{f}}\cr
&&
+ 4 D^{\overline{e}}F^{\overline{a} \underline{c} \underline{e}} F^{\overline{b}}{}_{\underline{c} \underline{f}} F_{\overline{e} \underline{e}}{}^{\underline{f}}
+ 2 F^{\overline{a} \underline{c} \underline{e}} F^{\overline{e}}{}_{\underline{c}}{}^{\underline{f}} F^{\overline{f}}{}_{\underline{e} \underline{f}} F^{\overline{b}}{}_{\overline{e} \overline{f}}
-2 F^{\overline{a} \underline{c} \underline{e}} F^{\overline{d}}{}_{\underline{e}}{}^{\underline{h}} F_{\overline{d} \underline{g} \underline{h}} F^{\overline{b}}{}_{\underline{c}}{}^{\underline{g}}  \cr
&& 
+ 2 F^{\overline{a} \underline{e} \underline{g}} F^{\overline{d}}{}_{\underline{e} \underline{f}} F_{\overline{d} \underline{g} \underline{h}} F^{\overline{b}\underline{f} \underline{h}} 
-  1/2 F^{\overline{a} \underline{e} \underline{f}} F^{\overline{d}}{}_{\underline{e} \underline{f}} F_{\overline{d} \underline{g} \underline{h}} F^{\overline{b} \underline{g} \underline{h}}
+ F^{\overline{e} \underline{c} \underline{d}} F^{\overline{f}}{}_{\underline{c} \underline{d}} F^{\overline{a}}{}_{\overline{e}}{}^{\underline{f}} F^{\overline{b}}{}_{\overline{f} \underline{f}} \left.\vphantom{2^{\frac12}}\right)\;,\cr&&\label{Lphi02}
\end{eqnarray} 
\begin{eqnarray}
 {\cal L}_{\Phi}^{(1,1)}&=&
F^{\underline{e}} \left(\vphantom{2^{\frac12}}\right. -D^{\underline{b}}D_{\underline{b}}F^{\overline{c} \underline{f} \underline{g}} F^{\overline{d}}{}_{\underline{f} \underline{g}} F_{\overline{c} \overline{d} \underline{e}}
-D^{\overline{d}}D^{\overline{e}}F^{\overline{f} \underline{d} \underline{c}} F_{\overline{d} \underline{d} \underline{c}} F_{\overline{e} \overline{f} \underline{e}}
-D^{\overline{c}}F_{\overline{c}}{}^{\underline{c} \underline{d}} D^{\overline{e}}F^{\overline{f}}{}_{\underline{c} \underline{d}} F_{\overline{e} \overline{f} \underline{e}}\cr
&&
-D^{\underline{h}}D_{\underline{e}}F^{\overline{c} \underline{f} \underline{g}} F^{\overline{d}}{}_{\underline{f} \underline{g}} F_{\overline{c} \overline{d} \underline{h}}  
 -D_{\underline{e}}F^{\overline{b} \underline{g} \underline{h}} D^{\underline{f}}F_{\overline{b}}{}^{\overline{d}}{}_{\underline{f}} F_{\overline{d} \underline{g} \underline{h}}
 -D^{\underline{b}}F^{\overline{b} \underline{g} \underline{h}} D_{\underline{b}}F_{\overline{b} \overline{d} \underline{e}}  F^{\overline{d}}{}_{ \underline{g} \underline{h}}\cr 
 && 
 -D_{\underline{e}}F^{\overline{c} \underline{g} \underline{f}} D^{\underline{h}}F^{\overline{d}}{}_{\underline{g} \underline{f}} F_{\overline{c} \overline{d} \underline{h}}
 -D^{\overline{d}}F^{\overline{e} \underline{d} \underline{f}} D^{\overline{f}}F_{\overline{d} \overline{e} \underline{e}} F_{\overline{f} \underline{d} \underline{f}} 
   -  \frac{1}{2} D^{\overline{e}}F^{\overline{f} \overline{g} \overline{h}} F_{\overline{e}}{}^{\underline{d} \underline{a}} F_{\overline{f} \underline{d} \underline{a}} F_{\overline{g} \overline{h} \underline{e}} \cr 
 && 
  -  \frac{1}{2} D^{\overline{c}}F^{\overline{f} \underline{d} \underline{c}} F_{\overline{c} \underline{d} \underline{c}} F^{\overline{g} \overline{h}}{}_{\underline{e}} F_{\overline{f} \overline{g} \overline{h}} 
  -  \frac{1}{2} D^{\overline{b}}F_{\overline{b}}{}^{\underline{d} \underline{c}} F^{\overline{f}}{}_{\underline{d} \underline{c}} F^{\overline{g} \overline{h}}{}_{\underline{e}} F_{\overline{f} \overline{g} \overline{h}} 
  -  \frac{1}{2} D^{\overline{d}}F^{\overline{f} \overline{g}}{}_{ \underline{e}}  F_{\overline{d}}{}^{\underline{b} \underline{f}} F^{\overline{h}}{}_{\underline{b} \underline{f}} F_{\overline{f} \overline{g} \overline{h}} \cr
 && 
 -2D^{\overline{c}}F^{\overline{e} \underline{d} \underline{f}} F_{\overline{c} \underline{d}}{}^{\underline{g}} F^{\overline{f}}{}_{\underline{f} \underline{g}} F_{\overline{e} \overline{f} \underline{e}} 
 + D^{\overline{b}}F_{\overline{b}}{}^{\underline{d} \underline{f}} F^{\overline{e}}{}_{\underline{d}}{}^{\underline{g}} F^{\overline{f}}{}_{\underline{f} \underline{g}} F_{\overline{e} \overline{f} \underline{e}}
 + D^{\overline{d}}F^{\overline{e} \overline{f}}{}_{\underline{e}}  F_{\overline{d}}{}^{\underline{b} \underline{g}} F_{\overline{e} \underline{b}}{}^{\underline{h}} F_{\overline{f} \underline{g} \underline{h}} 
  \cr
 &&
    + D^{\overline{d}}D^{\underline{d}}F^{\overline{e} \overline{f} \underline{c}} F_{\overline{d} \underline{d} \underline{c}} F_{\overline{e} \overline{f} \underline{e}}
    + D^{\underline{c}}F^{\overline{e} \overline{f} \underline{d}} D^{\overline{d}}F_{\overline{d} \underline{c} \underline{d}}  F_{\overline{e} \overline{f} \underline{e}} 
    + D^{\underline{d}}F^{\overline{d} \overline{e} \underline{f}} D^{\overline{f}}F_{\overline{d} \overline{e} \underline{e}}  F_{\overline{f} \underline{d} \underline{f}}  \cr 
  && 
 +  \frac{1}{2} D^{\overline{d}}F^{\overline{e} \overline{f} \underline{c}}  F_{\underline{c}}{}^{\underline{f} \underline{g}} F_{\overline{d} \underline{f} \underline{g}} F_{\overline{e} \overline{f} \underline{e}} 
 +  \frac{1}{2} D^{\overline{b}}F_{\overline{b}}{}^{\underline{d} \underline{c}}  F_{\underline{d} \underline{c}}{}^{\underline{g}} F^{\overline{e} \overline{f}}{}_{\underline{e}} F_{\overline{e} \overline{f} \underline{g}} 
  +  \frac{1}{2} D^{\overline{b}}F^{\underline{d} \underline{f} \underline{g}}  F_{\overline{b} \underline{d} \underline{f}} F^{\overline{e} \overline{f}}{}_{\underline{e}} F_{\overline{e} \overline{f} \underline{g}} \cr 
 &&  
 +  \frac{1}{2} D^{\overline{d}}F^{\overline{e} \overline{f}}{}_{ \underline{e}} F^{\underline{f} \underline{g} \underline{h}} F_{\overline{d} \underline{f} \underline{g}} F_{\overline{e} \overline{f} \underline{h}} 
  -2D^{\underline{b}}F^{\overline{c} \overline{e}}{}_{\underline{b}}  F_{\overline{c}}{}^{\underline{f} \underline{g}} F^{\overline{f}}{}_{\underline{f} \underline{g}} F_{\overline{e} \overline{f} \underline{e}} 
 + 2D^{\underline{g}}F^{\overline{c} \underline{d} \underline{f}} F^{\overline{e}}{}_{\underline{d} \underline{f}} F_{\overline{c}}{}^{\overline{f}}{}_{\underline{e}} F_{\overline{e} \overline{f} \underline{g}} \cr 
 && 
 + 2D^{\underline{g}}F^{\overline{e} \underline{d} \underline{f}} F^{\overline{c}}{}_{\underline{d} \underline{f}} F_{\overline{c}}{}^{\overline{f}}{}_{\underline{e}} F_{\overline{e} \overline{f} \underline{g}}
  -2D^{\underline{h}}F^{\overline{c} \overline{e}}{}_{\underline{e}}  F_{\overline{c}}{}^{\underline{f} \underline{g}} F^{\overline{f}}{}_{\underline{f} \underline{g}} F_{\overline{e} \overline{f} \underline{h}}
  + 2D^{\overline{d}}F^{\overline{e} \overline{g} \underline{c}}  F_{\overline{d} \underline{c}}{}^{\underline{b}} F_{\overline{e}}{}^{\overline{h}}{}_{\underline{e}} F_{\overline{g} \overline{h} \underline{b}}\cr
  &&
   -D^{\overline{b}}F_{\overline{b}}{}^{\underline{d} \underline{c}}  F^{\overline{e} \overline{g}}{}_{\underline{e}} F_{\overline{e}}{}^{\overline{h}}{}_{\underline{d}} F_{\overline{g} \overline{h} \underline{c}}
 -D^{\overline{d}}F^{\overline{e} \overline{g}}{}_{\underline{e}} F_{\overline{d}}{}^{\underline{b} \underline{f}} F_{\overline{e}}{}^{\overline{h}}{}_{\underline{b}} F_{\overline{g} \overline{h} \underline{f}}  
 \left. \vphantom{2^{\frac12}}\right)\cr 
 && 
-D^{\overline{d}}F^{\underline{a}} D^{\overline{e}}F^{\overline{f} \underline{d} \underline{e}} F_{\overline{d} \underline{d} \underline{e}} F_{\overline{e} \overline{f} \underline{a}}
 -D^{\underline{b}}F^{\underline{c}} D_{\underline{b}}F^{\overline{c} \underline{f} \underline{g}} F^{\overline{d}}{}_{\underline{f} \underline{g}} F_{\overline{c} \overline{d} \underline{c}}-D^{\underline{g}}F^{\underline{b}} D_{\underline{b}}F^{\overline{c} \underline{e} \underline{f}} F^{\overline{d}}{}_{\underline{e} \underline{f}} F_{\overline{c} \overline{d} \underline{g}}\cr
 &&
-  \frac{1}{2} D^{\overline{b}}F^{\underline{a}} F^{\overline{f}}{}_{\underline{d} \underline{e}} F_{\overline{b}}{}^{\underline{d} \underline{e}} F^{\overline{g} \overline{h}}{}_{\underline{a}} F_{\overline{f} \overline{g} \overline{h}}
+ D^{\overline{b}}F^{\underline{a}} F^{\overline{e}}{}_{\underline{d}}{}^{\underline{g}} F^{\overline{f}}{}_{\underline{f} \underline{g}} F_{\overline{b}}{}^{\underline{d} \underline{f}} F_{\overline{e} \overline{f} \underline{a}}
+ D^{\underline{d}}F^{\overline{e} \overline{f} \underline{e}} D^{\overline{d}}F^{\underline{c}} F_{\overline{d} \underline{d} \underline{e}} F_{\overline{e} \overline{f} \underline{c}}\cr 
&&   
 +  \frac{1}{2} D^{\overline{b}}F^{\underline{a}} F^{\underline{e} \underline{f} \underline{g}} F_{\overline{b} \underline{e} \underline{f}} F^{\overline{e} \overline{f}}{}_{\underline{a}} F_{\overline{e} \overline{f} \underline{g}} 
 + 2D^{\underline{g}}F^{\underline{b}} F^{\overline{c} \underline{e} \underline{f}} F^{\overline{e}}{}_{\underline{e} \underline{f}} F_{\overline{c}}{}^{\overline{f}}{}_{\underline{b}} F_{\overline{e} \overline{f} \underline{g}}
 -D^{\overline{b}}F^{\underline{a}} F_{\overline{b}}{}^{\underline{d} \underline{e}} F^{\overline{e} \overline{g}}{}_{\underline{a}} F_{\overline{e}}{}^{\overline{h}}{}_{\underline{d}} F_{\overline{g} \overline{h} \underline{e}}\cr 
 &&
 + D^{\underline{d}}F^{\overline{c} \underline{f} \underline{g}} F_{\underline{d}} F^{\underline{e}} F^{\overline{d}}{}_{\underline{f} \underline{g}} F_{\overline{c} \overline{d} \underline{e}}
 -F^{\underline{a}} F^{\underline{b}} F^{\overline{c} \underline{e} \underline{f}} F^{\overline{e}}{}_{\underline{e} \underline{f}} F_{\overline{c}}{}^{\overline{f}}{}_{\underline{a}} F_{\overline{e} \overline{f} \underline{b}} 
 + D^{\overline{d}}F^{\overline{e} \underline{d} \underline{e}} F^{\underline{c}} F^{\overline{c}} F_{\overline{c} \underline{d} \underline{e}} F_{\overline{d} \overline{e} \underline{c}} \cr 
 && +  \frac{1}{2} F^{\underline{a}} F^{\overline{a}} F_{\overline{a}}{}^{\underline{d} \underline{e}} F^{\overline{e}}{}_{\underline{d} \underline{e}} F^{\overline{f} \overline{g}}{}_{\underline{a}} F_{\overline{e} \overline{f} \overline{g}} 
 -F^{\underline{a}} F^{\overline{a}} F_{\overline{a}}{}^{\underline{d} \underline{f}} F^{\overline{e}}{}_{\underline{f} \underline{g}} F^{\overline{d}}{}_{\underline{d}}{}^{\underline{g}} F_{\overline{d} \overline{e} \underline{a}}\;.
 \label{Lphi11}
\end{eqnarray}

Dilaton fluxes appear on ${\cal L}^{(0)},\;{\cal L}^{(1)}\;,{\cal L}^{(2)}$ only up to quadratic order. This is not an accidental fact, but a general property that holds for all ${\cal L}^{(n)}$. Indeed, it is easily proven from two observations. First, the source of interactions is the quadratic action of the extended space which is quadratic in dilaton fluxes [see (\ref{CalR1}) or  (\ref{TildeCalR1})] and second, ${\cal F}_{\cal A B C}$ are independent of dilaton fluxes while ${\cal F}_{\cal A}$ are linear on dilaton fluxes of the double space $F_{a}$, as can be easily seen from the $O(D,D)$ flux decomposition in Appendix A of \cite{Baron:2020xel}. 

${\cal L}$ as it is presented here through (\ref{fullAction}) has the nice property of being exactly Lorentz invariant order by order; nevertheless, the price to pay for that is a huge number of terms in the action. Bianchi identities and section conditions can be, in principle, implemented to further simplify the Lagrangian. On the other hand, integration by parts or field redefinitions can be implemented too, but at the cost of realizing the Lorentz symmetry up to total derivative terms or at the cost of modifying the form of the $GGS$ transformations, respectively.

\section{Dilaton Flux cancellations in the $\alpha'$-expansion}

It is well known that the dilaton at tree level in String Theory enters in the action  either in the global overall factor $e^{-2\phi}$ or through the derivatives thereof. That is automatically satisfied by the $GBdRi$,  as the generalized dilaton on the extended space only appears in the overall factor $e^{-2d}\;(=\sqrt{|g|}\, e^{-2\phi})$ or through the dilaton fluxes. On the other hand, as we have commented above dilaton fluxes are present at all orders in $\alpha'$ but only in the form $(F_{A})^{n}$, with $n=0,1,2$. 
After moving to the SUGRA scheme derivatives of the dilaton will appear then up to $\partial_{m}\phi \partial_{n}\phi$ ,\footnote{That readily follows from equations (\ref{FAdecomp}) and (\ref{DilatonRedef}).} and therefore can be absorbed after integration by parts or by Lorentz covariant redefinitions of the metric and dilaton fields, except for the dilaton kinetic terms in ${\cal L}^{(0)}$. 

To see that, we first notice that terms linear in $\partial\phi$ can be rewritten without dilaton after integration by parts. Then, the only nontrivial couplings are those quadratic in derivatives of the dilaton, which in addition can be expressed as
\begin{eqnarray}
\sqrt{g} e^{-2\phi}\nabla_{m}\phi \nabla_{n}\phi\; T^{(N)}{}^{m n}&\equiv& \frac12 \sqrt{g} e^{-2\phi} \nabla_{m}\nabla_{n}\phi\; T^{(N)}{}^{m n} + \frac12 \sqrt{g} e^{-2\phi} \nabla_{n}\phi\; \nabla_{m}T^{(N)}{}^{m n}\cr 
&\equiv& \frac12 \sqrt{g} e^{-2\phi} \nabla_{m}\nabla_{n}\phi\; T^{(N)}{}^{m n} + \frac14 \sqrt{g} e^{-2\phi}\; \nabla_{n}\nabla_{m}T^{(N)}{}^{m n}\;,\cr&&\label{DpDp}
\end{eqnarray}
where ``$\equiv$" means up to total derivatives and $T^{(N)}{}^{mn}$ is an arbitrary (dilaton independent) tensor of Nth order in derivatives. 
We learn from (\ref{DpDp}) that terms proportional to $\partial_{m}\phi \partial_{n}\phi$ are equivalent to terms proportional to $\nabla_{m}\nabla_{n}\phi$, up to total derivatives and terms without derivatives of the dilaton.

We will now show that terms like these can be properly rewritten without dilatons after a suitable choice of variables.
 
 Indeed, after a field redefinition of the metric $g_{m n}\to g_{m n}+\delta g_{mn}$ and dilaton $\phi\to \phi+\delta\phi$, with $\delta g_{mn},\delta\phi\sim {\cal O}(\alpha')$ the second-derivative Lagrangian changes at linear order as 
${\cal L}^{(0)} +\delta{\cal L}^{(0)}$, with $\delta{\cal L}^{(0)}= G^{m n} \delta g_{m n} + \Phi \delta \phi $, where $G^{m n}$ and $\Phi$ refer to the $e.o.m.$ of the metric and dilaton, respectively. If we consider a field redefinition leaving $d$ invariant, the net effect is the addition of an extra piece at order ${\cal O}(\alpha')$: $\delta{\cal L}^{(0)}= \left(G^{m n}  + \frac14  g^{m n} \Phi \right) \delta g_{m n}= \left(R_{m n}  + 2 \nabla_{m}{\nabla_{n}\phi}-\frac14 H_{m p q} H_{n}{}^{p q} \right) \delta g^{m n}$. 

Hence, the choice $\delta g^{m n}=-\frac 14 T^{(2)}{}^{m n}$ cancels dilatons at order ${\cal O}(\alpha')$. Let us move to the following order in derivatives.

In this case, on top of the original couplings $\sim \nabla_{m}\nabla_{n}\phi\; T^{(4)}{}^{m n}$, there are some extra dilaton dependent terms at order ${\cal O}(\alpha'{}^2)$. Schematically from $\delta^{(2)}{\cal L}^{(0)}$ and $\delta{\cal L}^{(1)}$ due to nonlinear effects of the field redefinition. The crucial observation here is that these terms are still quadratic in dilaton derivatives because the field redefinitions (parameterized by $\delta g_{m n}$) are dilaton independent. Plugging all these interactions leads to a new term $ \nabla_{m}\nabla_{n}\phi\; \widetilde{T}^{(4)}{}^{m n}$ which again  can be eliminated by a new field redefinition parameterized by 
\begin{eqnarray}
\delta g^{m n}=-\frac 14 \widetilde{T}^{(4)}{}^{m n}\;\sim\; {\cal O}(\alpha'{}^2)
\end{eqnarray}

Clearly, we can proceed inductively to prove a very important lesson: there is a scheme in supergravity where the interactions related with the biparametric-DFT contain no dilaton beyond quadratic order, except for the exponential factor $e^{-2\phi}$. 
\medskip

Nevertheless, not necessarily does exist a frame in DFT where there is no dilaton flux beyond quadratic order. The reason is simple, on SUGRA we can only perform a change of basis by Lorentz covariant fields redefinitions (so that for instance the metric and dilaton remain being singlets). Similarly, the only allowed change of basis on DFT is due to duality covariant field redefinitions and therefore the previous results not necessarily extend to the DFT framework. 

The answer to this question is relevant as it could offer a simple scheme where the number of couplings in ${\cal L}$ is significantly reduced, but also to better understand the structure of the higher-derivative interactions in general and the role of the dilaton in particular when it is desired to linearly realize T-duality at the level of the Lagrangian. 

The first observation is that interactions being linear in dilaton fluxes can be moved from ${\cal R}_{\phi}$ into ${\cal R}_{\not\phi}$ after integration by parts. This follows easily from the identity
\begin{eqnarray}
e^{-2d} F_{A} T^{A} = e^{-2d} D_{A} T^{A} - \partial_{M}\left(e^{-2d} E^{M}{}_{A} T^{A}\right) \;.
\end{eqnarray}

Hence the only nontrivial couplings are those quadratic in dilaton fluxes or linear on derivatives thereof which in fact are equivalent, up to total derivatives. Indeed,
\begin{eqnarray}
e^{-2d} D_{A}F_{B} T^{A B} = e^{-2d} F_{A}F_{B} T^{A B} + \partial_{M}\left(e^{-2d} E^{M}{}_{A} F_{B} T^{A B}\right) \;.
\end{eqnarray}

Notice that antisymmetric derivatives ($D_{[A}F_{B]}$) are eliminated either after integration by parts or equivalently by using the Bianchi identity 
\begin{eqnarray}
D^{C} F_{C AB} &=& F^{C} F_{C AB}- 2 D_{[A}F_{B]}\;.
\end{eqnarray}
Regarding the quadratic terms, we must discuss separately the cases with mixed projections and those with equal projections. 

The former can be eliminated by field redefinitions as both $D_{\overline{a}}F_{\underline{b}}$ as well as $D_{\underline{a}}F_{\overline{b}}$ match the $e.o.m.$\footnote{After a field redefinition with $E^{M}{}_{A}\to E^{M}{}_{A}+\delta E^{M}{}_{A}$, $\delta E^{M}{}_{A} \sim{\cal O}(\alpha')$, the Lagrangian changes, at linear order, as ${\cal L}^{(0)}+{\cal G}_{[\overline{a} \underline{b}]} \delta E^{\overline{a} \underline{b}}$ where $\delta E^{A B}=\delta E_{M}{}^{A} E^{M B}$ and ${\cal G}_{[\overline{a} \underline{b}]}$ is the $e.o.m.$ of the generalized frame. The antisymmetry on the indices is due to the fact that the generalized frame is a constrained field, and it must hold $\delta E_{AB}=-\delta E_{BA}$. The components $\delta E_{\overline{a b}}$ and $\delta E_{\underline{a b}}$ vanish after use of Bianchi identities. } up to dilaton-independent or linear-dilaton terms. 
\begin{eqnarray}
{\cal G}^{[\overline{a}\underline{b}]}=D^{\overline{a}}F^{\underline{b}}+ (D_{\underline{c}}-F_{\underline{c}}) F^{\overline{a}\underline{b c}}- F^{\underline{d}\overline{c a}} F_{\overline{c}\underline{d}}{}^{\underline{b}}\;,\label{eom}
\end{eqnarray}
or equivalently 
\begin{eqnarray}
{\cal G}^{[\overline{a}\underline{b}]}&=&2\,D^{\underline{b}}F^{\overline{a}}-2 (D_{\overline{c}}-F_{\overline{c}}) F^{\overline{a}\underline{b}\overline{c}}-2 F^{\underline{d}\overline{c a}}F_{\overline{c}\underline{d}}{}^{\underline{b}}\;.
\end{eqnarray}

We are left with equal projection terms $D_{\overline{a}}F_{\overline{b}}$ and  $D_{\underline{a}}F_{\underline{b}}$. 

These are much more subtle. It turns out that those terms appearing in the original expression in (\ref{fullAction}) with such a structure can be absorbed at a given order but at a very high cost as it requires a duality-covariant (dilaton-dependent)\footnote{This must be contrasted with the SUGRA scheme where $\delta g_{mn}$ was dilaton-independent.} field redefinition of the generalized vielbein. Then, it reintroduces dilaton dependence at the next order in derivatives, and it can be shown that it occurs in a way that cannot be reabsorbed any more. But more importantly, once we study the nonlinear effects of the (dilaton-dependent) field redefinitions, dilaton fluxes appear at any order, not only quadratically as in the original expansion. 

For instance let us consider ${\cal L}^{(1)}_a$
\begin{eqnarray}
{\cal L}^{(1)}_{a}
 &\supset&2\;D_{\overline{a}}\left(F_{\overline{b}} F^{\overline{a} \underline{c} \underline{d}} F^{\overline{b}}{}_{\underline{c} \underline{d}} \right)\,- \, F_{\overline{a}} F_{\overline{b}} F^{\overline{a} \underline{c} \underline{d}} F^{\overline{b}}{}_{\underline{c} \underline{d}} \cr
&=& 
 F_{\overline{a}} F_{\overline{b}} F^{\overline{a} \underline{c} \underline{d}} F^{\overline{b}}{}_{\underline{c} \underline{d}}  
 +2 (D_{\overline{a}}-F_{\overline{a}})\left(F_{\overline{b}} F^{\overline{a} \underline{c} \underline{d}} F^{\overline{b}}{}_{\underline{c} \underline{d}} \right)\;,
\label{L1-DilatonTerms}
 \end{eqnarray}
where the last term in the second line is a total derivative (after multiplication by $e^{-2d}$). The first term can be alternatively rewritten, up to total-derivative and dilaton-independent couplings as $D_{\overline{a}} F_{\overline{b}} F^{\overline{a} \underline{c} \underline{d}} F^{\overline{b}}{}_{\underline{c} \underline{d}}$ but it leads to a dead end as there are no identities for $D_{(\overline{a}} F_{\overline{b})}$. On the other hand, the first term in (\ref{L1-DilatonTerms}) can be worked out by the use of Bianchi identities as
 \begin{eqnarray}
 F_{\overline{a}} F_{\overline{b}} F^{\overline{a} \underline{c} \underline{d}} F^{\overline{b}}{}_{\underline{c} \underline{d}} &=&
  F_{\overline{a}} F_{B} F^{\overline{a} \underline{c} \underline{d}} F^{B}{}_{\underline{c} \underline{d}} -
   F_{\overline{a}} F_{\underline{b}} F^{\overline{a} \underline{c} \underline{d}} F^{\underline{b}}{}_{\underline{c} \underline{d}} \cr
  &=& F_{\overline{a}} F^{\overline{a} \underline{c} \underline{d}} \left(2D_{\underline{c}}F_{\underline{d}}+D^{B}F_{B\underline{c} \underline{d}}\right) -
D_{\underline{b}}\left(F_{\overline{a}}  F^{\overline{a} \underline{c} \underline{d}} F^{\underline{b}}{}_{\underline{c} \underline{d}}\right)\cr
&+&
\left(D_{\underline{b}}-F_{\underline{b}}\right)\left(F_{\overline{a}}  F^{\overline{a} \underline{c} \underline{d}} F^{\underline{b}}{}_{\underline{c} \underline{d}}\right)\cr
  &=& 2D_{\underline{c}}F_{\underline{d}} F_{\overline{a}} F^{\overline{a} \underline{c} \underline{d}}
 - 
 D_{\underline{b}}F_{\overline{a}} \; F^{\overline{a} \underline{c} \underline{d}} F^{\underline{b}}{}_{\underline{c} \underline{d}}
 +
  D_{\overline{a}} \left( F^{\overline{a} \underline{c} \underline{d}} D^{\overline{b}}F_{\overline{b} \underline{c} \underline{d}}
-D_{\underline{b}} F^{\overline{a} \underline{c} \underline{d}} \; F^{\underline{b}}{}_{\underline{c} \underline{d}} \right)
\cr
&+&
\left(D_{\underline{a}}-F_{\underline{a}}\right)\left(F_{\overline{b}}  F^{\overline{b} \underline{c} \underline{d}} F^{\underline{a}}{}_{\underline{c} \underline{d}}\right)
- \left(D_{\overline{a}}-F_{\overline{a}}\right) \left( F^{\overline{a} \underline{c} \underline{d}} D^{\overline{b}}F_{\overline{b} \underline{c} \underline{d}}
-D_{\underline{b}} F^{\overline{a} \underline{c} \underline{d}} \; F^{\underline{b}}{}_{\underline{c} \underline{d}} \right)\;.\label{FaFb1}
\cr && 
\end{eqnarray}
The last line is a total derivative, the last term in the first line is dilaton independent while the second term can be rewritten as 
\begin{eqnarray}
D_{\underline{b}}F_{\overline{a}}  F^{\overline{a} \underline{c} \underline{d}} F^{\underline{b}}{}_{\underline{c} \underline{d}}&=&
\frac12{\cal G}_{[\overline{a} \underline{b}]} F^{\overline{a} \underline{c} \underline{d}} F^{\underline{b}}{}_{\underline{c} \underline{d}}
-F_{\overline{a}\underline{b}\overline{e}} D^{\overline e}\left( F^{\overline{a} \underline{c} \underline{d}} F^{\underline{b}}{}_{\underline{c} \underline{d}}\right)
+
 F_{\underline{e}\overline{f}\overline{a}} F^{\overline{f}\underline{e}}{}_{\underline{b}} F^{\overline{a} \underline{c} \underline{d}} F^{\underline{b}}{}_{\underline{c} \underline{d}}\cr
 &+&
\left(D_{\overline a}- F_{\overline a}\right)\left(F^{\overline{e}\underline{b}\overline{a}} F_{\overline{e} \underline{c} \underline{d}} F_{\underline{b}}{}^{\underline{c} \underline{d}}  \right)\;. \label{FaFb2}
\end{eqnarray}
The first term is proportional to the $e.o.m.$ and so can be absorbed in a (duality-covariant) field redefinition, the second and third ones are dilaton independent and the last one is once again a total derivative.

Regarding the first term in the last equality of (\ref{FaFb1}), it also can be worked out in order to eliminate the dilaton but at the cost of implementing a dilaton-dependent field redefinition, indeed
 \begin{eqnarray}
 2\,D_{\underline{c}}F_{\underline{d}} F_{\overline{a}} F^{\overline{a} \underline{c} \underline{d}}
&=&
 \left({\cal G}_{[\overline{a} \underline{b}]} +2\left(D^{\overline e}- F^{\overline e}\right)F_{\overline{a}\underline{b}\overline{e}} + 2F_{\underline{e}\overline{f}\overline{a}} F^{\overline{f}\underline{e}}{}_{\underline{b}}\right)
 \left(D_{\underline{c}}-F_{\underline{c}}\right)F^{\overline{a} \underline{b} \underline{c}} 
+\, D_{\overline{a}} \left(F_{\underline{c} \underline{d}}{}^{B} D_{B}F^{\overline{a} \underline{c} \underline{d}}\right) \cr
&+&
\, \left(D_{\overline{a}}-F_{\overline{a}}\right) \left(- F_{\underline{c} \underline{d}}{}^{B} D_{B}F^{\overline{a} \underline{c} \underline{d}}\right)
+\left(D_{\underline{a}}-F_{\underline{a}}\right)
\left(-2\,F_{\overline{c}} \left(D_{\underline{b}}-F_{\underline{b}}\right)F^{\overline{c} \underline{a} \underline{b}} \right)\cr
&=&
{\cal G}_{[\overline{a} \underline{b}]} 
 \left(D_{\underline{c}}-F_{\underline{c}}\right)F^{\overline{a} \underline{b} \underline{c}} 
-2\, D_{\underline{a}}
F_{\overline e} \;  F^{\overline{c} \underline{a} \underline{b}} F_{\overline{c}\underline{b}}{}^{\overline{e}} 
-2\, F_{\overline e}   F^{\overline{c} \underline{a} \underline{b}} 
D_{\underline{a}} F_{\overline{c}\underline{b}}{}^{\overline{e}} 
\cr
 &+&
 2\, F^{\overline{c} \underline{a} \underline{b}} D_{\underline{a}}
\left(D_{\overline e} F_{\overline{c}\underline{b}}{}^{\overline{e}} 
+ F_{\underline{e}\overline{f}\overline{c}} F^{\overline{f}\underline{e}}{}_{\underline{b}}\right)
+ D_{\overline{a}} \left(F_{\underline{c} \underline{d}}{}^{B} D_{B}F^{\overline{a} \underline{c} \underline{d}}\right) \cr
&-& 2 \left(D_{\underline{a}}-F_{\underline{a}}\right)\left[F^{\overline{c} \underline{a} \underline{b}} 
\left(\left(D^{\overline e}- F^{\overline e}\right)F_{\overline{c}\underline{b}\overline{e}} + F_{\underline{e}\overline{f}\overline{c}} F^{\overline{f}\underline{e}}{}_{\underline{b}}\right)
+F_{\overline{c}} \left(D_{\underline{b}}-F_{\underline{b}}\right)F^{\overline{c} \underline{a} \underline{b}} \right]
\cr
 &+&
\, \left(D_{\overline{a}}-F_{\overline{a}}\right) \left(- F_{\underline{c} \underline{d}}{}^{B} D_{B}F^{\overline{a} \underline{c} \underline{d}}\right)\cr
&=&
 \, 2  F_{\overline{e}\underline{b}\overline{a}} D^{\overline a}\left(    F^{\overline{c} \underline{b} \underline{d}} F_{\overline{c}\underline{d}}{}^{\overline{e}}    \right)
-
2F_{\underline{d}\overline{f}\overline{a}} F^{\overline{f}\underline{d}}{}_{\underline{b}}
 F^{\overline{c} \underline{b} \underline{e}} F_{\overline{c}\underline{e}}{}^{\overline{a}} 
 -2\, D_{\overline a}  \left( F^{\overline{c} \underline{d} \underline{b}} 
D_{\underline{d}} F_{\overline{c}\underline{b}}{}^{\overline{a}}\right) 
\cr
 &+&
 2\, F^{\overline{c} \underline{a} \underline{b}} D_{\underline{a}}
\left(D_{\overline e} F_{\overline{c}\underline{b}}{}^{\overline{e}} 
+ F_{\underline{e}\overline{f}\overline{c}} F^{\overline{f}\underline{e}}{}_{\underline{b}}\right)
+ D_{\overline{a}} \left(F_{\underline{c} \underline{d}}{}^{B} D_{B}F^{\overline{a} \underline{c} \underline{d}}\right)\cr
&+&
{\cal G}_{[\overline{a} \underline{b}]} \left[
 \left(D_{\underline{c}}-F_{\underline{c}}\right)F^{\overline{a} \underline{b} \underline{c}} - F^{\overline{c} \underline{b} \underline{e}} F_{\overline{c}\underline{e}}{}^{\overline{a}} \right]
 \cr
&-& 2 \left(D_{\underline{a}}-F_{\underline{a}}\right)\left[
\left(\left(D^{\overline e}- F^{\overline e}\right)F_{\overline{c}\underline{b}\overline{e}} + F_{\underline{e}\overline{f}\overline{c}} F^{\overline{f}\underline{e}}{}_{\underline{b}}\right)
F^{\overline{c} \underline{a} \underline{b}} 
+F_{\overline{c}} \left(D_{\underline{b}}-F_{\underline{b}}\right)F^{\overline{c} \underline{a} \underline{b}} \right]
\cr
 &+&
\left(D_{\overline a} -  F_{\overline a} \right)  \left[2 \; F^{\overline{c} \underline{d} \underline{b}} 
D_{\underline{d}} F_{\overline{c}\underline{b}}{}^{\overline{a}} 
- F_{\underline{c} \underline{d}}{}^{B} D_{B}F^{\overline{a} \underline{c} \underline{d}}
-2 \; F_{\overline{e}\underline{b}}{}^{\overline{a}}  F^{\overline{c} \underline{b} \underline{d}} F_{\overline{c}\underline{d} }{}^{\overline{e}} \right]\;,
\label{FaFb3}
\end{eqnarray}
with the first two lines being dilaton independent, the third line can be absorbed in a dilaton-dependent field redefinition and the last two lines are total derivatives. Plugging (\ref{FaFb1}), (\ref{FaFb2}) and (\ref{FaFb3}) we indeed verify that dilaton fluxes can be completely eliminated from ${\cal L}^{(1)}$. Some comments are in order
\begin{enumerate}[i]
    \item It can be shown by a careful inspection of the first term in (\ref{L1-DilatonTerms}) that there is no other way to eliminate the dilaton. 
    \item Even though we managed to eliminate dilaton fluxes there is not a real simplification in ${\cal L}^{(1)}$ as we end up with more terms in the Lagrangian than the original ones.
    \item We got rid of dilaton fluxes at order ${\cal O}(\alpha')$, but the necessary dilaton-dependent field redefinitions reintroduce dilaton fluxes at subleading orders and they appear now at any power, not just quadratically.
    \item Similar steps as those followed here can be implemented to cancel the dilaton couplings appearing on ${\cal L}^{(2)}$ in (\ref{Lphi02}) and (\ref{Lphi11})\footnote{Indeed, it is also possible to cancel all the original terms containing dilatons at arbitrary order by implementing dilaton field redefinitions directly in the extended space.}. Nevertheless, it can be shown that the dilaton interactions at ${\cal O}(\alpha'{}^2)$ with origin in the field redefinitions at order ${\cal O}(\alpha')$ cannot be completely eliminated by implementing field redefinitions, integration by parts and use of Bianchi identities or section conditions. 
    \item The absence of a dilaton-free frame in DFT is not in conflict with SUGRA. What we state here is that the dilaton cannot be eliminated by duality-covariant field redefinitions, the situation drastically changes if we move to the SUGRA variables which indeed require duality noncovariant field redefinitions.
\end{enumerate}

All the previous statements have tedious and technical proofs, but only the last one is in some way interesting and have useful implications and so we will focus on this one on the next section.

\section{Perturbative field-redefinitions for the metric and dilaton fields}
The $GBdRi$ is naturally established in the framework of Double Field Theory; however, in order to make contact with supergravity it is required to find appropriate noncovariant field redefinitions. This is the only step which is noniterative and must be performed by brute force, $i.e.$ by demanding the metric and dilaton (as well as the two-form, in the bosonic case: $a=b$) to be invariant under Lorentz transformations. In this section we will show that the field redefinition for both the metric and the dilaton can be fixed by demanding the derivatives of the dilaton cancel out when we move from DFT to SUGRA variables.
\medskip

The relevant piece of the action of the extended space is 
\begin{eqnarray}
e^{-2d} {\cal R}_1 = e^{-2d}\left[2 \left({\cal D}_{\overline{\cal A}} {\cal F}^{\overline{\cal A}} - {\cal D}_{\underline{\cal A}} {\cal F}^{\underline{\cal A}} \right)- \left({\cal F}_{\overline{\cal A}} {\cal F}^{\overline{\cal A}} - {\cal F}_{\underline{\cal A}} {\cal F}^{\underline{\cal A}} \right)\right]\;,\label{CalR1}
\end{eqnarray}
or equivalently (up to total derivative terms)
\begin{eqnarray}
e^{-2d} \tilde{\cal R}_1 = e^{-2d} \left({\cal F}_{\overline{\cal A}} {\cal F}^{\overline{\cal A}} - {\cal F}_{\underline{\cal A}} {\cal F}^{\underline{\cal A}} \right)\;.\label{TildeCalR1}
\end{eqnarray}
The $O(D,D)$ decomposition of the dilaton fluxes is displayed in Appendix A of \cite{Baron:2020xel}. Nevertheless, these expressions admit a further simplification after which the ${\cal F}_{\cal A}$ components reduce to
\begin{eqnarray}
{\cal F}{}_{\underline{a}}&=&
\left(F_{\underline{b}} -\partial_{\underline{b}}\right) (\chi^{\frac12}){}_{\underline{a}}{}^{\underline{b}} \;,\cr\cr
{\cal F}{}_{{\underline\alpha}}&=&
 {\underline e}^{{\ut\mu}}{}_{{\underline\alpha}} \, \left[\vphantom{\frac12}\right.
\left(F_{\underline{b}}-\partial_{\underline{b}}\right)   \left( \Omega^{{\tilde\nu}}{}_{{\ut\mu}}   {\cal E}_{{\tilde\nu}}{}^{\underline{b}} \right)
 -\, \left(F_{\overline{b}}-\partial_{\overline{b}}\right) \left( (\Pi^{\frac12})^{{\ut\nu}}{}_{{\ut\mu}}   {\cal E}_{{\ut\nu}}{}^{\overline{b}} \right)
\left. \vphantom{\frac12}\right] \;.
\end{eqnarray}
Considering the other components, one arrives at
\begin{eqnarray}
{\cal F}_{\overline{\cal A}}{\cal F}^{\overline{\cal A}} - {\cal F}_{\underline{\cal A}}{\cal F}^{\underline{\cal A}}&=& 
F_{\overline{a}} F^{\overline{a}} - F_{\underline{a}} F^{\underline{a}}\cr 
&-&
2\; \Pi^{{\ut\mu}}{}^{{\ut\nu}}  {\cal E}_{{\ut\mu}}{}^{\overline{a}}  {\cal E}_{{\ut\nu}}{}^{\overline{b}} 
F_{\overline{a}} F_{\overline{b}} 
+2\; \Pi^{{\tilde\nu}}{}^{{\tilde\rho}}   {\cal E}_{{\tilde\nu}}{}^{\underline{a}}   {\cal E}_{{\tilde\rho}}{}^{\underline{b}} 
  F_{\underline{a}}  F_{\underline{b}} 
  +4  \left( \Omega^{{\tilde\nu}}{}_{{\ut\mu}}   {\cal E}_{{\tilde\nu}}{}^{\underline{a}} \right)
 \left( (\Pi^{\frac12})^{{\ut\rho}}{}^{{\ut\mu}}   {\cal E}_{{\ut\rho}}{}^{\overline{b}} \right)F_{\underline{a}} F_{\overline{b}}\cr 
 &-&
 2\; \left(F_{\overline{a}}-\partial_{\overline{a}}\right) \left\{\vphantom{2^{frac12}}\right. (\chi^{\frac12})^{\overline{a}}{}^{\overline{c}}
 \partial^{\overline{b}}(\chi^{\frac12}){}_{\overline{c}}{}_{\overline{b}}
 + \Omega_{{\tilde\mu}}{}^{{\ut\nu}}  {\cal E}_{{\ut\nu}}{}^{\overline{a}}  \left[
    \partial_{\overline{b}} \left( \Omega^{{\tilde\mu}}{}^{{\ut\rho}}   {\cal E}_{{\ut\rho}}{}^{\overline{b}} \right)
 + \partial_{\underline{b}} \left( (\Pi^{\frac12})^{{\tilde\rho}}{}^{{\tilde\mu}}   {\cal E}_{{\tilde\rho}}{}^{\underline{b}} \right)
  \right]\cr 
&&\;\;\;\;\;\;\;\;\;\;\;\;\;\;\;\;\;\;\;\;\; 
 + \left( (\Pi^{\frac12})^{{\ut\nu}}{}^{{\ut\mu}}   {\cal E}_{{\ut\nu}}{}^{\overline{a}} \right) \left[
\partial_{\underline{b}} \left( \Omega^{{\tilde\rho}}{}_{{\ut\mu}}   {\cal E}_{{\tilde\rho}}{}^{\underline{b}} \right)
\left.\vphantom{2^{frac12}}- 
  \partial_{\overline{b}} \left( (\Pi^{\frac12})^{{\ut\rho}}{}_{{\ut\mu}}   {\cal E}_{{\ut\rho}}{}^{\overline{b}} \right) 
  \right]
  \right\}
 \cr 
 &+&
 2\; \left(F_{\underline{a}}-\partial_{\underline{a}}\right)  \left\{\vphantom{2^{frac12}}\right. 
  (\chi^{\frac12})^{\underline{a}}{}^{\underline{c}}
 \partial^{\underline{b}}(\chi^{\frac12}){}_{\underline{c}}{}_{\underline{b}}
 + \left( \Omega^{{\tilde\nu}}{}_{{\ut\mu}}   {\cal E}_{{\tilde\nu}}{}^{\underline{a}} \right)\left[
    \partial_{\underline{b}} \left( \Omega^{{\tilde\rho}}{}^{{\ut\mu}}   {\cal E}_{{\tilde\rho}}{}^{\underline{b}} \right)
 - \partial_{\overline{b}} \left( (\Pi^{\frac12})^{{\ut\rho}}{}^{{\ut\mu}}   {\cal E}_{{\ut\rho}}{}^{\overline{b}} \right) 
\right]\cr
&&\;\;\;\;\;\;\;\;\;\;\;\;\;\;\;\;\;\;\;\;\; 
- \left( (\Pi^{\frac12})^{{\tilde\nu}}{}^{{\tilde\mu}}   {\cal E}_{{\tilde\nu}}{}^{\underline{a}} \right)  \left[
  \partial_{\underline{b}} \left( (\Pi^{\frac12})^{{\tilde\rho}}{}_{{\tilde\mu}}   {\cal E}_{{\tilde\rho}}{}^{\underline{b}} \right)
  - \partial_{\overline{b}} \left( \Omega_{{\tilde\mu}}{}^{{\ut\rho}}   {\cal E}_{{\ut\rho}}{}^{\overline{b}} \right)\right]
\left.\vphantom{2^{frac12}} \right\}
 \cr 
 &-&
 2\; \partial_{\overline{a}}\left\{ \vphantom{2^{frac12}} \right.(\chi^{\frac12})^{\overline{a}}{}^{\overline{c}}
 \partial^{\overline{b}}(\chi^{\frac12}){}_{\overline{c}}{}_{\overline{b}}
 + \left( \Omega_{{\tilde\mu}}{}^{{\ut\nu}}  {\cal E}_{{\ut\nu}}{}^{\overline{a}} \right) \left[
    \partial_{\overline{b}} \left( \Omega^{{\tilde\mu}}{}^{{\ut\rho}}   {\cal E}_{{\ut\rho}}{}^{\overline{b}} \right)
 + 
\partial_{\underline{b}} \left( (\Pi^{\frac12})^{{\tilde\rho}}{}^{{\tilde\mu}}   {\cal E}_{{\tilde\rho}}{}^{\underline{b}} \right) \right]
\cr 
&&\;\;\;\;\;\;\;\;\;\;\;\;\;\;\;\;\;\;\;\;\; 
- \left( (\Pi^{\frac12})^{{\ut\nu}}{}^{{\ut\mu}}   {\cal E}_{{\ut\nu}}{}^{\overline{a}} \right)  \left[
  \partial_{\overline{b}} \left( (\Pi^{\frac12})^{{\ut\rho}}{}_{{\ut\mu}}   {\cal E}_{{\ut\rho}}{}^{\overline{b}} \right)
  - 
\partial_{\underline{b}} \left( \Omega^{{\tilde\rho}}{}_{{\ut\mu}}   {\cal E}_{{\tilde\rho}}{}^{\underline{b}} \right)\right]
\left.\vphantom{2^{frac12}} \right\}
\cr 
&+& 
\left\{ \vphantom{2^{frac12}} \right. \partial_{\overline{a}}(\chi^{\frac12}){}^{\overline{a}}{}^{\overline{c}} \partial^{\overline{b}}(\chi^{\frac12}){}_{\overline{c}}{}_{\overline{b}} 
 +  \partial_{\overline{a}} \left( \Omega_{{\tilde\mu}}{}^{{\ut\nu}}   {\cal E}_{{\ut\nu}}{}^{\overline{a}} \right)\left[
 \partial_{\overline{b}} \left( \Omega^{{\tilde\mu}}{}^{{\ut\rho}}   {\cal E}_{{\ut\rho}}{}^{\overline{b}} \right) 
 - 
\partial_{\underline{b}}\left( (\Pi^{\frac12})^{{\tilde\rho}}{}^{{\tilde\mu}}   {\cal E}_{{\tilde\rho}}{}^{\underline{b}} \right)\right]
\cr 
  &&
 \;\;\;\;\;\;\;\;\;\;\;\;\;\;\;\;\;\;\;\;\; - 
 \partial_{\overline{a}} \left( (\Pi^{\frac12})^{{\ut\nu}}{}^{{\ut\mu}}   {\cal E}_{{\ut\nu}}{}^{\overline{a}} \right)\left[ \partial_{\overline{b}}\left( (\Pi^{\frac12})^{{\ut\rho}}{}_{{\ut\mu}}   {\cal E}_{{\ut\rho}}{}^{\overline{b}} \right)
 - \partial_{\underline{b}}  \left( \Omega^{{\tilde\rho}}{}_{{\ut\mu}}   {\cal E}_{{\tilde\rho}}{}^{\underline{b}} \right)\right]
\left.\vphantom{2^{frac12}} \right\}\cr
&+&
2\; \partial_{\underline{a}}\left\{ \vphantom{2^{frac12}} \right.(\chi^{\frac12})^{\underline{a}}{}^{\underline{c}}
 \partial^{\underline{b}}(\chi^{\frac12}){}_{\underline{c}}{}_{\underline{b}}
 + \left( \Omega^{{\tilde\nu}}{}_{{\ut\mu}}   {\cal E}_{{\tilde\nu}}{}^{\underline{a}} \right)\left[
    \partial_{\underline{b}} \left( \Omega^{{\tilde\rho}}{}^{{\ut\mu}}   {\cal E}_{{\tilde\rho}}{}^{\underline{b}} \right)
 -  
\partial_{\overline{b}} \left( (\Pi^{\frac12})^{{\ut\rho}}{}^{{\ut\mu}}   {\cal E}_{{\ut\rho}}{}^{\overline{b}} \right) \right] \cr 
 &&
  \;\;\;\;\;\;\;\;\;\;\;\;\;\;\;\;\;\;\;\;\; - \left( (\Pi^{\frac12})^{{\tilde\nu}}{}^{{\tilde\mu}}   {\cal E}_{{\tilde\nu}}{}^{\underline{a}} \right)  \left[
  \partial_{\underline{b}} \left( (\Pi^{\frac12})^{{\tilde\rho}}{}_{{\tilde\mu}}   {\cal E}_{{\tilde\rho}}{}^{\underline{b}} \right)
  + 
\partial_{\overline{b}} \left( \Omega_{{\tilde\mu}}{}^{{\ut\rho}}   {\cal E}_{{\ut\rho}}{}^{\overline{b}} \right) \right]
\left.\vphantom{2^{frac12}} \right\}\cr 
 &-&
\left\{ \vphantom{2^{frac12}} \right. \partial_{\underline{a}}(\chi^{\frac12}){}^{\underline{c}}{}^{\underline{a}} 
\partial^{\underline{b}}(\chi^{\frac12}){}_{\underline{c}}{}_{\underline{b}}
 +  \partial_{\underline{a}} \left( \Omega^{{\tilde\nu}}{}_{{\ut\mu}}   {\cal E}_{{\tilde\nu}}{}^{\underline{a}} \right)\left[
 \partial_{\underline{b}} \left( \Omega^{{\tilde\rho}}{}^{{\ut\mu}}   {\cal E}_{{\tilde\rho}}{}^{\underline{b}} \right) 
 - 
\partial_{\overline{b}}\left( (\Pi^{\frac12})^{{\ut\rho}}{}^{{\ut\mu}}   {\cal E}_{{\ut\rho}}{}^{\overline{b}} \right)\right]\cr 
  &&
  \;\;\;\;\;\;\;\;\;\;\;\;\;\;\;\;\;\;\;\;\; 
 - \partial_{\underline{a}} \left( (\Pi^{\frac12})^{{\tilde\rho}}{}^{{\tilde\mu}}   {\cal E}_{{\tilde\rho}}{}^{\underline{a}} \right) \left[
  \partial_{\underline{b}}\left( (\Pi^{\frac12})^{{\tilde\nu}}{}_{{\tilde\mu}}   {\cal E}_{{\tilde\nu}}{}^{\underline{b}} \right)
 + 
 \partial_{\overline{b}}  \left( \Omega_{{\tilde\mu}}{}^{{\ut\nu}}   {\cal E}_{{\ut\nu}}{}^{\overline{b}} \right)\right]
\left.\vphantom{2^{frac12}} \right\}
\;.\cr&&
\label{DilatonAction}\;\;\;\;
\end{eqnarray}
The first line gives the second-derivative interactions of the dilaton, $\tilde{R}_1$ in (\ref{tildeR1}). It is not difficult to see that the full $R_1$ in (\ref{R1}) can be generated if we bypass the integration by parts in the extended space and perform the flux decomposition in (\ref{CalR1}) instead of (\ref{TildeCalR1}).
The second line gives (implicitly) the higher-derivative interactions of the dilaton field in the DFT framework, the third to sixth lines are total derivatives and the remaining terms are dilaton independent. 

The second line explicitly shows that there are no dilaton derivatives beyond quadratic order as $\Pi, {\cal E}$, and $\Omega$ are dilaton independent. The last term in the second line can be absorbed {\it a priori} in a field redefinition as it leads, after integration by parts, to 
\begin{eqnarray}
&& 4  \left( \Omega^{{\tilde\nu}}{}_{{\ut\mu}}   {\cal E}_{{\tilde\nu}}{}^{\underline{a}} \right)
 \left( (\Pi^{\frac12})^{{\ut\rho}}{}^{{\ut\mu}}   {\cal E}_{{\ut\rho}}{}^{\overline{b}} \right)F_{\underline{a}} F_{\overline{b}}
 \;=\;\;
 4\, \left[ F^{\overline{c}\underline{d}}{}_{\underline{a}} F_{\underline{d}\overline{c b}} 
 - F_{\overline{b}\underline{a}\overline{c}}
D^{\overline{c}}
 + D_{\overline{b}} D_{\underline{a}}\right] \left( \Omega^{{\tilde\nu}}{}_{{\ut\mu}}   {\cal E}_{{\tilde\nu}}{}^{\underline{a}} \right)
 \left( (\Pi^{\frac12})^{{\ut\rho}}{}^{{\ut\mu}}   {\cal E}_{{\ut\rho}}{}^{\overline{b}} \right)
  \;
 \cr&&
 \;\;\;\;\;\;\;\;\;\;\; \;\;\;\;\;\;\;\;\;\,
 +\;2\,{\cal G}_{[\overline{b}\underline{a}]} \left( \Omega^{{\tilde\nu}}{}_{{\ut\mu}}   {\cal E}_{{\tilde\nu}}{}^{\underline{a}} \right)
 \left( (\Pi^{\frac12})^{{\ut\rho}}{}^{{\ut\mu}}   {\cal E}_{{\ut\rho}}{}^{\overline{b}} \right)
  + 4 (D^{\overline{c}}-F^{\overline{c}})\left( F_{\overline{b}\underline{a}\overline{c}}
 \left( \Omega^{{\tilde\nu}}{}_{{\ut\mu}}   {\cal E}_{{\tilde\nu}}{}^{\underline{a}} \right)
 \left( (\Pi^{\frac12})^{{\ut\rho}}{}^{{\ut\mu}}   {\cal E}_{{\ut\rho}}{}^{\overline{b}} \right)\right) \cr&&
 \;\;\;\;\;\;\;
  -4 (D_{\overline{b}}- F_{\overline{b}}) D_{\underline{a}}\left(\left( \Omega^{{\tilde\nu}}{}_{{\ut\mu}}   {\cal E}_{{\tilde\nu}}{}^{\underline{a}} \right)
 \left( (\Pi^{\frac12})^{{\ut\rho}}{}^{{\ut\mu}}   {\cal E}_{{\ut\rho}}{}^{\overline{b}} \right)\right)
- 4 (D_{\underline{a}} - F_{\underline{a}})\left( F_{\overline{b}} \left( \Omega^{{\tilde\nu}}{}_{{\ut\mu}}   {\cal E}_{{\tilde\nu}}{}^{\underline{a}} \right)
 \left( (\Pi^{\frac12})^{{\ut\rho}}{}^{{\ut\mu}}   {\cal E}_{{\ut\rho}}{}^{\overline{b}} \right)\right)\,.\cr&&
\end{eqnarray}

The $r.h.s.$ on the first line is dilaton independent, the first term in the second line can be absorbed in a (dilaton-independent) field redefinition of the generalized vielbein and the last three are total derivatives.

The first and second terms on the second line of (\ref{DilatonAction}) cannot be eliminated with duality-covariant (and dilaton-independent) field redefinitions.

\subsection{No dilaton ansatz and the field redefinitions}
Let us consider once again (\ref{DilatonAction}), but this time let us perform the GL(D) decomposition of the dilaton fluxes.
\begin{eqnarray}
{\cal F}_{\overline{\cal A}}{\cal F}^{\overline{\cal A}} - {\cal F}_{\underline{\cal A}}{\cal F}^{\underline{\cal A}}&\supset& 
\left(\eta^{\overline{a} \overline{b}} -2\; \Pi^{{\ut\mu}}{}^{{\ut\nu}}  {\cal E}_{{\ut\mu}}{}^{\overline{a}}  {\cal E}_{{\ut\nu}}{}^{\overline{b}} 
 \right)F_{\overline{a}} F_{\overline{b}} 
- \left(\eta^{\underline{a} \underline{b}} - 2\; \Pi^{{\tilde\nu}}{}^{{\tilde\rho}}   {\cal E}_{{\tilde\nu}}{}^{\underline{a}}   {\cal E}_{{\tilde\rho}}{}^{\underline{b}} 
 \right) F_{\underline{a}} F_{\underline{b}}\cr &&
  +4  \left( \Omega^{{\tilde\nu}}{}_{{\ut\mu}}   {\cal E}_{{\tilde\nu}}{}^{\underline{a}} \right)
 \left( (\Pi^{\frac12})^{{\ut\rho}}{}^{{\ut\mu}}   {\cal E}_{{\ut\rho}}{}^{\overline{b}} \right)F_{\underline{a}} F_{\overline{b}}\cr
 &\supset&
 2 \left[\vphantom{2^{\frac12}}\right.
 \left(g^{\overline{a} \overline{b}} -2\; \Pi^{{\ut\rho}}{}^{{\ut\sigma}}  {\cal E}_{{\ut\rho}}{}^{\overline{a}}  {\cal E}_{{\ut\sigma}}{}^{\overline{b}} 
 \right) \overline{e}^{m}{}_{\overline{a}} \overline{e}^{n}{}_{\overline{b}} 
- \left(- g^{\underline{a} \underline{b}} - 2\; \Pi^{{\tilde\rho}}{}^{{\tilde\sigma}}   {\cal E}_{{\tilde\rho}}{}^{\underline{a}}   {\cal E}_{{\tilde\sigma}}{}^{\underline{b}} 
 \right)\overline{e}^{m}{}_{\underline{a}} \overline{e}^{n}{}_{\underline{b}} \cr &&
  +4  \left( \Omega^{{\tilde\rho}}{}_{{\ut\sigma}}   {\cal E}_{{\tilde\rho}}{}^{\underline{a}} \right)
 \left( (\Pi^{\frac12})^{{\ut\lambda}}{}^{{\ut\sigma}} {\cal E}_{{\ut\lambda}}{}^{\overline{b}} \right)\overline{e}^{m}{}_{\underline{a}} \overline{e}^{n}{}_{\overline{b}} 
 \left.\vphantom{2^{\frac12}}\right] \partial_{m}{\overline\phi} \partial_{n}{\overline\phi}
\;\cr
 &\supset&
 2 \left[\vphantom{2^{\frac12}}\right.
 \left(g^{\overline{a} \overline{b}} -2\; \Pi^{{\ut\rho}}{}^{{\ut\sigma}}  {\cal E}_{{\ut\rho}}{}^{\overline{a}}  {\cal E}_{{\ut\sigma}}{}^{\overline{b}}
 \right) \overline{e}^{m}{}_{\overline{a}} \overline{e}^{n}{}_{\overline{b}} 
+ \left(g^{\underline{a} \underline{b}} + 2\; \Pi^{{\tilde\rho}}{}^{{\tilde\sigma}}   {\cal E}_{{\tilde\rho}}{}^{\underline{a}}   {\cal E}_{{\tilde\sigma}}{}^{\underline{b}} 
 \right)\overline{e}^{m}{}_{\underline{a}} \overline{e}^{n}{}_{\underline{b}} \cr &&
  +4  \left( \Omega^{{\tilde\rho}}{}_{{\ut\sigma}}   {\cal E}_{{\tilde\rho}}{}^{\underline{a}} \right)
 \left( (\Pi^{\frac12})^{{\ut\lambda}}{}^{{\ut\sigma}} {\cal E}_{{\ut\lambda}}{}^{\overline{b}} \right)\overline{e}^{m}{}_{\underline{a}} \overline{e}^{n}{}_{\overline{b}} 
 \left.\vphantom{2^{\frac12}}\right] \partial_{m}{\phi} \partial_{n}{\phi}\;,\label{dphidphi}
\end{eqnarray}
where in the second inclusion we have used\footnote{It readily follows from the frame decomposition
\begin{eqnarray}
E_{M}{}^{A}=\frac{1}{\sqrt{2}}\left(\begin{matrix}
\overline{e}^{m \underline{a}}
&
\overline{e}^{m \overline{a}} 
\cr
\left(\overline{b}_{m n} - \overline{g}_{m n}\right) \overline{e}^{n \underline{a}} 
&
\left(\overline{b}_{m n} + \overline{g}_{m n}\right) \overline{e}^{n \overline{a}}
\end{matrix} \right)\;,
\;\;\;\;\;\;\;\;\;\;\;
E^{M}{}_{A}=\eta^{M N}\,E_{N}{}^{B}\, \eta_{B A}\,,
\end{eqnarray}
and the invariant metrics
\begin{eqnarray}
\eta^{M N}=\left(\begin{matrix} 0 & \delta_{m}{}^{n}
\cr
\delta^{m}{}_{n}&0
\end{matrix} \right) \;\;,\;\;\;\;\;\;\;\;\;\;
\eta_{A B}=\left(\begin{matrix}
-\overline{g}_{\underline{a b}}&0
\cr
0&\overline{g}_{\overline{a b}}
\end{matrix} \right) \;.
\end{eqnarray}
}
\begin{eqnarray}
F_{\overline{a}} &=& \sqrt 2 \overline{e}^{m}{}_{\overline{a}} \partial_{m}\overline{\phi} 
+\frac{1}{\sqrt 2}  \overline{\omega}_{\overline{b a}}{}^{\overline b} - 
\left(\partial^{\widetilde m}-terms\right)\;,\cr
F_{\underline{a}} &=& - \sqrt 2 \overline{e}^{m}{}_{\underline{a}} \partial_{m}\overline{\phi} 
-\frac{1}{\sqrt 2} \overline{\omega}_{\underline{b a}}{}^{\underline b} 
+ \left(\partial^{\widetilde m}-terms\right)\;,\label{FAdecomp}
\end{eqnarray}
$\overline{\omega}$ in the first and second line denotes the spin connection computed with the pair of independent vielbeins $\overline{e}^{m}{}_{\overline a}$ and $\overline{e}^{m}{}_{\underline a}$ respectively. The line over the fields is here to distinguish them from the physical (Lorentz singlets) variables.\footnote{$\partial^{\widetilde m}$-terms vanish if we solve the strong constraint with the standard supergravity section condition.} Finally, in the last inclusion we have used the relation 
\begin{eqnarray}
\overline\phi=\phi+\frac14 ln\left|\overline{g}/g\right|\;,\;\;\;\;\;\;\;\;\;
g=det\left(g_{mn}\right)\label{DilatonRedef}
\end{eqnarray}
which is a consequence of the fact that the generalized dilaton $d$ does not receive corrections neither under Lorentz nor under $O(D,D)$.

The requirement of no derivative of dilaton at higher orders leads to the identification of the last two lines in (\ref{dphidphi}) with the kinetic term of the dilaton: $4\, g^{m n} \partial_{m}{\phi} \partial_{n}{\phi}$, which implies
\begin{eqnarray}
g^{m n}=
\overline{g}^{m n} 
+  \Pi^{{\tilde\rho}}{}^{{\tilde\sigma}} {\cal E}_{{\tilde\rho}}{}^{\underline{a}} {\cal E}_{{\tilde\sigma}}{}^{\underline{b}} 
\overline{e}^{m}{}_{\underline{a}} \overline{e}^{n}{}_{\underline{b}}
- \Pi^{{\ut\rho}}{}^{{\ut\sigma}} {\cal E}_{{\ut\rho}}{}^{\overline{a}} {\cal E}_{{\ut\sigma}}{}^{\overline{b}} 
 \overline{e}^{m}{}_{\overline{a}} \overline{e}^{n}{}_{\overline{b}} 
+2 \left( \Omega^{{\tilde\rho}}{}_{{\ut\sigma}} {\cal E}_{{\tilde\rho}}{}^{\underline{a}} \right)
 \left( (\Pi^{\frac12})^{{\ut\lambda}}{}^{{\ut\sigma}} {\cal E}_{{\ut\lambda}}{}^{\overline{b}} \right)\overline{e}^{(m}{}_{\underline{a}} \overline{e}^{n)}{}_{\overline{b}} \;\;.\;\;\;\;\cr&&\label{gmapBP}
\end{eqnarray}

It is worth mentioning that exactly the same conclusion is obtained if we avoid the integration by parts. This time the flux decomposition is performed on ${\cal R}_1$ instead of on $\tilde{\cal R}_1$ and the metric ansatz is obtained after comparison with $-4\partial^{m}\phi \partial_{m}\phi+4\nabla^{m}\nabla_{m}\phi$ whose difference with $+4\partial^{m}\phi \partial_{m}\phi$ is a total derivative.

Let us introduce $\Lambda_{\overline{a}}{}^{\underline {b}}=\overline{e}^{m}{}_{\overline{a}}\, \overline{e}_{m}{}^{\underline{b}}\in O(1,D-1)$, then the physical and duality covariant frames are related, up to an arbitrary Lorentz transformation $O^{\underline a}{}_{b}$, via 
\begin{eqnarray}
e_{m}{}^{a}&=&  \overline{e}_{m}{}^{\underline b}(\Xi^{-\frac12})_{\underline b}{}^{\underline c} O_{\underline{c}}{}^{a}\;,
\end{eqnarray}
with
\begin{eqnarray}
\Xi_{\underline a}{}^{\underline b}&=& 
\delta_{\underline{a}}{}^{\underline{b}}
+  \Pi^{{\tilde\rho}}{}^{{\tilde\sigma}} {\cal E}_{{\tilde\rho}}{}_{\underline{a}} {\cal E}_{{\tilde\sigma}}{}^{\underline{b}} 
- \Pi^{{\ut\rho}}{}^{{\ut\sigma}} {\cal E}_{{\ut\rho}}{}^{\overline{c}} {\cal E}_{{\ut\sigma}}{}^{\overline{d}} \Lambda_{\overline{c}}{}_{\underline {a}} \Lambda_{\overline{d}}{}^{\underline {b}}
+2 \left( \Omega^{{\tilde\rho}}{}_{{\ut\sigma}} {\cal E}_{{\tilde\rho}}{}_{(\underline{a}} \right)
 \left( (\Pi^{\frac12})^{{\ut\lambda}}{}^{{\ut\sigma}} {\cal E}_{{\ut\lambda}}{}^{\overline{c}} \right) \Lambda_{|\overline{c }|\underline{d})} \overline{g}^{\underline{d b}}\;,
\cr&&
\end{eqnarray}
so that 
\begin{equation}
\phi=\overline\phi-\frac14 ln \left[\vphantom{2^{\frac12}} det\left(\delta_{\underline{a}}{}^{\underline{b}}
+  \Pi^{{\tilde\rho}}{}^{{\tilde\sigma}} {\cal E}_{{\tilde\rho}}{}_{\underline{a}} {\cal E}_{{\tilde\sigma}}{}^{\underline{b}} 
- \Pi^{{\ut\rho}}{}^{{\ut\sigma}} {\cal E}_{{\ut\rho}}{}^{\overline{c}} {\cal E}_{{\ut\sigma}}{}^{\overline{d}} \Lambda_{\overline{c}}{}_{\underline {a}} \Lambda_{\overline{d}}{}^{\underline {b}}
+ 2  \left( \Omega^{{\tilde\rho}}{}_{{\ut\sigma}} {\cal E}_{{\tilde\rho}}{}_{(\underline{a}} \right)
 \left( (\Pi^{\frac12})^{{\ut\lambda}}{}^{{\ut\sigma}} {\cal E}_{{\ut\lambda}}{}^{\overline{c}} \right) \Lambda_{|\overline{c }|\underline{d})} \overline{g}^{\underline{d b}}\right)\right]\;.\;\nonumber
 \end{equation}\begin{equation}\label{phimapBP}
\end{equation}

These expressions simplify considerably in the monoparametric case (heterotic DFT) with, for instance $a=0$, as there is no scalars ($\Omega=0$), hence
\begin{eqnarray}
g^{m n}=\overline{g}^{m n} 
+  {\cal E}_{{\tilde\mu}}{}^{\underline{a}} {\cal E}^{{\tilde\mu}}{}^{\underline{b}}
\overline{e}^{m}{}_{\underline{a}} \overline{e}^{n}{}_{\underline{b}}\;,\label{gmapMP}
\end{eqnarray}
\begin{eqnarray}
\phi=\overline\phi-\frac14 ln \left[\vphantom{2^{\frac12}} det\left(\delta_{\underline{b}}{}^{\underline{a}}+ {\cal E}^{{\tilde\mu}}{}_{\underline{b}} {\cal E}_{{\tilde\mu}}{}^{\underline{a}}\right)\right]\;.\label{phimapMP}
\end{eqnarray}

Unfortunately, this argument cannot be used to fix the field redefinitions of the two-form and therefore we still need to compute it by hand.

In order to test the proposals we can work out (\ref{gmapBP}) up to ${\cal O}(\alpha'{}^2)$ and compare with the expressions displayed in \cite{Baron:2020xel}. Actually, we do not need to compute it here as we can read it directly from ${\cal L}$. Once again we follow the shorter route, we integrate by parts on (\ref{R1}), (\ref{L1}), (\ref{Lphi02}) and (\ref{Lphi11}) to accommodate all dilaton fluxes, up to ${\cal O}(\alpha'{}^2)$, on
\begin{eqnarray}
{\cal L}&\supset& F_{\overline a}F^{\overline a} - F_{\underline a}F^{\underline a}\; 
+\; a\; F_{\overline{a}} F_{\overline{b}} F^{\overline{a} \underline{a} \underline{b}} F^{\overline{b}}{}_{\underline{a} \underline{b}} 
+\; b \; F_{\underline{a}} F_{\underline{b}} F^{\underline{a} \overline{a} \overline{b}} F^{\underline{b}}{}_{\overline{a} \overline{b}}\cr
&+& a^2 \;F_{\overline{a}}F_{\overline{b}} \left[\vphantom{2^{\frac12}}\right. - D^{\underline{c}}F^{\overline{a} \underline{e} \underline{f}} D_{\underline{c}}F^{\overline{b}}{}_{\underline{e} \underline{f}} 
-4 D^{\overline{e}}F^{\overline{b} \underline{c} \underline{e}} F^{\overline{a}}{}_{\underline{c}\underline{f}} F_{\overline{e} \underline{e}}{}^{\underline{f}}
+ 2 D^{\overline{b}}F^{\overline{e} \underline{c} \underline{e}} F^{\overline{a}}{}_{\underline{c} \underline{f}} F_{\overline{e} \underline{e}}{}^{\underline{f}} 
\cr
&&
+ 2 D^{\underline{f}}F^{\overline{b} \underline{d} \underline{e}} F^{\overline{f}}{}_{\underline{d} \underline{e}} F^{\overline{a}}{}_{\overline{f} \underline{f}} 
 - F_{\overline{e}}{}^{\underline{c} \underline{d}} F^{\overline{g}}{}_{\underline{c} \underline{d}} F^{\overline{a}}{}_{\overline{g} \underline{f}} F^{\overline{b} \overline{e}}{}^{\underline{f}} 
 + 2 F^{\overline{a} \underline{c}}{}^{\underline{g}} F^{\overline{e}}{}_{\underline{e}}{}^{\underline{h}} F_{\overline{e} \underline{g} \underline{h}} F^{\overline{b}}{}_{\underline{c}}{}^{\underline{e}}\cr
&&
 - 2 F^{\overline{d}}{}_{\underline{c}}{}^{\underline{g}} F_{\overline{d} \underline{e}}{}^{\underline{h}} F^{\overline{a}}{}_{\underline{g} \underline{h}} F^{\overline{b} \underline{c} \underline{e}} 
+ 1/2 F^{\overline{e}}{}_{\underline{c} \underline{d}} F^{\overline{a} \underline{g} \underline{h}} F_{\overline{e} \underline{g} \underline{h}} F^{\overline{b}\underline{c} \underline{d}}
-2 F^{\overline{f}}{}_{\underline{c}}{}^{\underline{f}} F^{\overline{g}}{}_{\underline{e} \underline{f}} F^{\overline{b} \underline{c} \underline{e}} F^{\overline{a}}{}_{\overline{f} \overline{g}} \left.\vphantom{2^{\frac12}}\right]\cr
&-& b^2 \;F_{\underline{a}}F_{\underline{b}} \left[\vphantom{2^{\frac12}}\right. - D^{\overline{c}}F^{\underline{a} \overline{e} \overline{f}} D_{\overline{c}}F^{\underline{b}}{}_{\overline{e} \overline{f}} 
-4 D^{\underline{e}}F^{\underline{b} \overline{c} \overline{e}} F^{\underline{a}}{}_{\overline{c}\overline{f}} F_{\underline{e} \overline{e}}{}^{\overline{f}}
+ 2 D^{\underline{b}}F^{\underline{e} \overline{c} \overline{e}} F^{\underline{a}}{}_{\overline{c} \overline{f}} F_{\underline{e} \overline{e}}{}^{\overline{f}} 
\cr
&&
+ 2 D^{\overline{f}}F^{\underline{b} \overline{d} \overline{e}} F^{\underline{f}}{}_{\overline{d} \overline{e}} F^{\underline{a}}{}_{\underline{f} \overline{f}} 
 - F_{\underline{e}}{}^{\overline{c} \overline{d}} F^{\underline{g}}{}_{\overline{c} \overline{d}} F^{\underline{a}}{}_{\underline{g} \overline{f}} F^{\underline{b} \underline{e}}{}^{\overline{f}} 
 + 2 F^{\underline{a} \overline{c}}{}^{\overline{g}} F^{\underline{e}}{}_{\overline{e}}{}^{\overline{h}} F_{\underline{e} \overline{g} \overline{h}} F^{\underline{b}}{}_{\overline{c}}{}^{\overline{e}}\cr
&&
 - 2 F^{\underline{d}}{}_{\overline{c}}{}^{\overline{g}} F_{\underline{d} \overline{e}}{}^{\overline{h}} F^{\underline{a}}{}_{\overline{g} \overline{h}} F^{\underline{b} \overline{c} \overline{e}} 
+ 1/2 F^{\underline{e}}{}_{\overline{c} \overline{d}} F^{\underline{a} \overline{g} \overline{h}} F_{\underline{e} \overline{g} \overline{h}} F^{\underline{b}\overline{c} \overline{d}}
-2 F^{\underline{f}}{}_{\overline{c}}{}^{\overline{f}} F^{\underline{g}}{}_{\overline{e} \overline{f}} F^{\underline{b} \overline{c} \overline{e}} F^{\underline{a}}{}_{\underline{f} \underline{g}} \left.\vphantom{2^{\frac12}}\right]\cr
&+& \; a b \left[\vphantom{2^{\frac12}}\right. F^{\overline{a}}F^{\underline{b}} \left(\vphantom{2^{\frac12}}\right. 
 D^{\underline{d}}F^{\overline{e} \overline{f} \underline{e}}F_{\overline{a} \underline{d} \underline{e}} F_{\overline{e} \overline{f} \underline{b}}
+  \frac{1}{2}  F^{\underline{e} \underline{f} \underline{g}} F_{\overline{a} \underline{e} \underline{f}} F^{\overline{e} \overline{f}}{}_{\underline{b}} F_{\overline{e} \overline{f} \underline{g}}
- F_{\overline{a}}{}^{\underline{d} \underline{e}} F^{\overline{e} \overline{g}}{}_{\underline{b}} F_{\overline{e}}{}^{\overline{h}}{}_{\underline{d}} F_{\overline{g} \overline{h} \underline{e}}\cr
&&
 -D^{\overline{d}}F^{\underline{e} \underline{f} \overline{e}}F_{\underline{b} \overline{d} \overline{e}} F_{\underline{e} \underline{f} \overline{a}}
-  \frac{1}{2}  F^{\overline{e} \overline{f} \overline{g}} F_{\underline{b} \overline{e} \overline{f}} F^{\underline{e} \underline{f}}{}_{\overline{a}} F_{\underline{e} \underline{f} \overline{g}}
+ F_{\underline{b}}{}^{\overline{d} \overline{e}} F^{\underline{e} \underline{g}}{}_{\overline{a}} F_{\underline{e}}{}^{\underline{h}}{}_{\overline{d}} F_{\underline{g} \underline{h} \overline{e}} \left.\vphantom{2^{\frac12}}\right)
 \cr 
 &&
 +F^{\overline{a}}F^{\overline{b}}\left(\vphantom{2^{\frac12}}\right.  D_{\overline{b}}F^{\underline{c} \overline{e} \overline{f}} F^{\underline{d}}{}_{\overline{e} \overline{f}} F_{\underline{c} \underline{d} \overline{a}}
 -  F^{\underline{c} \overline{e} \overline{f}} F^{\underline{e}}{}_{\overline{e} \overline{f}} F_{\underline{c}}{}^{\underline{f}}{}_{\overline{b}} F_{\underline{e} \underline{f} \overline{a}} \left. \vphantom{2^{\frac12}}\right)
\cr&&
 -F^{\underline{a}}F^{\underline{b}}\left(\vphantom{2^{\frac12}}\right.  D_{\underline{b}}F^{\overline{c} \underline{e} \underline{f}} F^{\overline{d}}{}_{\underline{e} \underline{f}} F_{\overline{c} \overline{d} \underline{a}}
 -  F^{\overline{c} \underline{e} \underline{f}} F^{\overline{e}}{}_{\underline{e} \underline{f}} F_{\overline{c}}{}^{\overline{f}}{}_{\underline{b}} F_{\overline{e} \overline{f} \underline{a}} \left. \vphantom{2^{\frac12}}\right)
\left.\vphantom{2^{\frac12}}\right]\;,
\end{eqnarray}
from which we read
\begin{eqnarray}
g^{m n} &=& \overline{g}^{m n} 
+\; \frac{a}{2} \overline{e}^{m}{}_{\overline{a}} \overline{e}^{n}{}_{\overline{b}}  F^{\overline{a} \underline{a} \underline{b}} F^{\overline{b}}{}_{\underline{a} \underline{b}} +\;\frac{b}{2}\; \overline{e}^{m}{}_{\underline{a}} \overline{e}^{n}{}_{\underline{b}}  F^{\underline{a} \overline{a} \overline{b}} F^{\underline{b}}{}_{\overline{a} \overline{b}}
\cr
&+& 
\;\frac{a^2}{2}\;\overline{e}^{m}{}_{\overline{a}} \overline{e}^{n}{}_{\overline{b}}
\;\left(\vphantom{2^{\frac12}}\right. - D^{\underline{c}}F^{\overline{a} \underline{e} \underline{f}} D_{\underline{c}}F^{\overline{b}}{}_{\underline{e} \underline{f}} 
-4 D^{\overline{e}}F^{\overline{b} \underline{c} \underline{e}} F^{\overline{a}}{}_{\underline{c}\underline{f}} F_{\overline{e} \underline{e}}{}^{\underline{f}}
+ 2 D^{\overline{b}}F^{\overline{e} \underline{c} \underline{e}} F^{\overline{a}}{}_{\underline{c} \underline{f}} F_{\overline{e} \underline{e}}{}^{\underline{f}} 
\cr
&&
+ 2 D^{\underline{f}}F^{\overline{b} \underline{d} \underline{e}} F^{\overline{f}}{}_{\underline{d} \underline{e}} F^{\overline{a}}{}_{\overline{f} \underline{f}} 
 - F_{\overline{e}}{}^{\underline{c} \underline{d}} F^{\overline{g}}{}_{\underline{c} \underline{d}} F^{\overline{a}}{}_{\overline{g} \underline{f}} F^{\overline{b} \overline{e}}{}^{\underline{f}} 
 + 2 F^{\overline{a} \underline{c}}{}^{\underline{g}} F^{\overline{e}}{}_{\underline{e}}{}^{\underline{h}} F_{\overline{e} \underline{g} \underline{h}} F^{\overline{b}}{}_{\underline{c}}{}^{\underline{e}}\cr
&&
 - 2 F^{\overline{d}}{}_{\underline{c}}{}^{\underline{g}} F_{\overline{d} \underline{e}}{}^{\underline{h}} F^{\overline{a}}{}_{\underline{g} \underline{h}} F^{\overline{b} \underline{c} \underline{e}} 
+ 1/2 F^{\overline{e}}{}_{\underline{c} \underline{d}} F^{\overline{a} \underline{g} \underline{h}} F_{\overline{e} \underline{g} \underline{h}} F^{\overline{b}\underline{c} \underline{d}}
-2 F^{\overline{f}}{}_{\underline{c}}{}^{\underline{f}} F^{\overline{g}}{}_{\underline{e} \underline{f}} F^{\overline{b} \underline{c} \underline{e}} F^{\overline{a}}{}_{\overline{f} \overline{g}} \left.\vphantom{2^{\frac12}}\right)
\cr
&-& 
\frac{b^2}{2}\; \overline{e}^{m}{}_{\underline{a}} \overline{e}^{n}{}_{\underline{b}} \left(\vphantom{2^{\frac12}}\right. - D^{\overline{c}}F^{\underline{a} \overline{e} \overline{f}} D_{\overline{c}}F^{\underline{b}}{}_{\overline{e} \overline{f}} 
-4 D^{\underline{e}}F^{\underline{b} \overline{c} \overline{e}} F^{\underline{a}}{}_{\overline{c}\overline{f}} F_{\underline{e} \overline{e}}{}^{\overline{f}}
+ 2 D^{\underline{b}}F^{\underline{e} \overline{c} \overline{e}} F^{\underline{a}}{}_{\overline{c} \overline{f}} F_{\underline{e} \overline{e}}{}^{\overline{f}} \cr
&&
+ 2 D^{\overline{f}}F^{\underline{b} \overline{d} \overline{e}} F^{\underline{f}}{}_{\overline{d} \overline{e}} F^{\underline{a}}{}_{\underline{f} \overline{f}} 
 - F_{\underline{e}}{}^{\overline{c} \overline{d}} F^{\underline{g}}{}_{\overline{c} \overline{d}} F^{\underline{a}}{}_{\underline{g} \overline{f}} F^{\underline{b} \underline{e}}{}^{\overline{f}} 
 + 2 F^{\underline{a} \overline{c}}{}^{\overline{g}} F^{\underline{e}}{}_{\overline{e}}{}^{\overline{h}} F_{\underline{e} \overline{g} \overline{h}} F^{\underline{b}}{}_{\overline{c}}{}^{\overline{e}}\cr
&&
 - 2 F^{\underline{d}}{}_{\overline{c}}{}^{\overline{g}} F_{\underline{d} \overline{e}}{}^{\overline{h}} F^{\underline{a}}{}_{\overline{g} \overline{h}} F^{\underline{b} \overline{c} \overline{e}} 
+ 1/2 F^{\underline{e}}{}_{\overline{c} \overline{d}} F^{\underline{a} \overline{g} \overline{h}} F_{\underline{e} \overline{g} \overline{h}} F^{\underline{b}\overline{c} \overline{d}}
-2 F^{\underline{f}}{}_{\overline{c}}{}^{\overline{f}} F^{\underline{g}}{}_{\overline{e} \overline{f}} F^{\underline{b} \overline{c} \overline{e}} F^{\underline{a}}{}_{\underline{f} \underline{g}} \left.\vphantom{2^{\frac12}}\right)
\cr
&-& \frac{a b}{2}\; \left(\vphantom{{\frac12}^{\frac12}}\right. - \overline{e}^{m}{}_{\overline{a}} \overline{e}^{n}{}_{\underline{b}}  \left(\vphantom{2^{\frac12}}\right. 
 D^{\underline{d}}F^{\overline{e} \overline{f} \underline{e}}F_{\overline{a} \underline{d} \underline{e}} F_{\overline{e} \overline{f} \underline{b}}
+  \frac{1}{2}  F^{\underline{e} \underline{f} \underline{g}} F_{\overline{a} \underline{e} \underline{f}} F^{\overline{e} \overline{f}}{}_{\underline{b}} F_{\overline{e} \overline{f} \underline{g}}
- F_{\overline{a}}{}^{\underline{d} \underline{e}} F^{\overline{e} \overline{g}}{}_{\underline{b}} F_{\overline{e}}{}^{\overline{h}}{}_{\underline{d}} F_{\overline{g} \overline{h} \underline{e}}\cr
&&
 -D^{\overline{d}}F^{\underline{e} \underline{f} \overline{e}}F_{\underline{b} \overline{d} \overline{e}} F_{\underline{e} \underline{f} \overline{a}}
-  \frac{1}{2}  F^{\overline{e} \overline{f} \overline{g}} F_{\underline{b} \overline{e} \overline{f}} F^{\underline{e} \underline{f}}{}_{\overline{a}} F_{\underline{e} \underline{f} \overline{g}}
+ F_{\underline{b}}{}^{\overline{d} \overline{e}} F^{\underline{e} \underline{g}}{}_{\overline{a}} F_{\underline{e}}{}^{\underline{h}}{}_{\overline{d}} F_{\underline{g} \underline{h} \overline{e}} \left.\vphantom{2^{\frac12}}\right)
 \cr 
 &&
 +\overline{e}^{m}{}_{\overline{a}} \overline{e}^{n}{}_{\overline{b}} \left(\vphantom{2^{\frac12}}\right.  D_{\overline{b}}F^{\underline{c} \overline{e} \overline{f}} F^{\underline{d}}{}_{\overline{e} \overline{f}} F_{\underline{c} \underline{d} \overline{a}}
 -  F^{\underline{c} \overline{e} \overline{f}} F^{\underline{e}}{}_{\overline{e} \overline{f}} F_{\underline{c}}{}^{\underline{f}}{}_{\overline{b}} F_{\underline{e} \underline{f} \overline{a}} \left. \vphantom{2^{\frac12}}\right)
\cr&&
 -\overline{e}^{m}{}_{\underline{a}} \overline{e}^{n}{}_{\underline{b}} \left(\vphantom{2^{\frac12}}\right.  D_{\underline{b}}F^{\overline{c} \underline{e} \underline{f}} F^{\overline{d}}{}_{\underline{e} \underline{f}} F_{\overline{c} \overline{d} \underline{a}}
 -  F^{\overline{c} \underline{e} \underline{f}} F^{\overline{e}}{}_{\underline{e} \underline{f}} F_{\overline{c}}{}^{\overline{f}}{}_{\underline{b}} F_{\overline{e} \overline{f} \underline{a}} \left. \vphantom{2^{\frac12}}\right)
  \left.\vphantom{{\frac12}^{\frac12}}\right)\;.
\end{eqnarray}

This field redefinition is somehow more general than the one displayed in \cite{Baron:2020xel}, as here we still preserve the full double Lorentz group. To make contact with $loc.\; cit.$ we partially fix the Lorentz group by imposing $\overline{e}^{m}{}_{\overline a}= \overline{e}^{m}{}_{b}\, \delta^{b}{}_{\overline a}$ and $\overline{e}^{m}{}_{\underline a}= \overline{e}^{m}{}_{b}\, \delta^{b}{}_{\underline a}$, which in addition implies $D_{\overline a} = - \frac{1}{\sqrt2}\delta_{\overline a}{}^{b} \partial_{b}$ y $D_{\underline a} =  \frac{1}{\sqrt2}\delta_{\underline a}{}^{b} \partial_{b}$ , with $\partial_{b}= \overline{e}^{m}{}_{b}\,\partial_{m}$ as well as
\begin{eqnarray}
F_{\overline{a}\underline{b c}} &=& \frac{1}{\sqrt{2}} \delta^{a}_{\overline{a}} \delta^{b}_{\underline{b}} \delta^{c}_{\underline{c}} \overline{\omega}^{(+)}{}_{a b c}
\;,\;\;\;\;\;\;\;\;\;\;  
F_{\overline{a b c}} = -  \frac{1}{\sqrt{2}} \delta^{a}_{\overline{a}} \delta^{b}_{\overline{b}} \delta^{c}_{\overline{c}} \left(2\overline{\omega}^{(+)}{}_{[a b c]} +  \overline{\omega}^{(-)}{}_{[a b c]}\right)\;,\cr
F_{\underline{a}\overline{b c}} &=&  \frac{1}{\sqrt{2}} \delta^{a}_{\underline{a}} \delta^{b}_{\overline{b}} \delta^{c}_{\overline{c}} \overline{\omega}^{(-)}{}_{a b c}
\;,\;\;\;\;\;\;\;\;\;\; 
F_{\underline{a b c}} = -  \frac{1}{\sqrt{2}} \delta^{a}_{\underline{a}} \delta^{b}_{\underline{b}} \delta^{c}_{\underline{c}} \left(2\overline{\omega}^{(-)}{}_{[a b c]} +  \overline{\omega}^{(+)}{}_{[a b c]}\right)\;,
\end{eqnarray}
where the torsionful spin connections are given by 
\begin{eqnarray}
\overline{\omega}^{(\pm)}_{a b c}=\overline{e}^{m}{}_{a}\left( \overline{\omega}_{m b c}\pm \frac12 \overline{H}_{m n p} \overline{e}^{n}{}_{b} \overline{e}^{p}{}_{c}\right)\,,
\end{eqnarray}
where the line over the spin connection and the three-form means these are computed in terms of the Lorentz noncovariant fields $\overline{e}_{m}{}^{a}$ and $\overline{b}_{m n}$. Hence,
\begin{eqnarray}
g^{m n} &=& \overline{g}^{m n} 
+\; \frac{a}{4} \overline{\omega}^{(-)}{}^{m b c} \overline{\omega}^{(-)}{}^{n}{}_{b c} +\;\frac{b}{4}\; \overline{\omega}^{(+)}{}^{m b c} \overline{\omega}^{(+)}{}^{n}{}_{b c}
\cr
&+& 
\frac{a^2}{8}\overline{e}^{(m}{}_{a} \overline{e}^{n)}{}_{b}
\;\left[\vphantom{2^{\frac12}}\right.  \partial^{c} \overline{\omega}^{(-)}{}^{a e f} \partial_{c}\overline{\omega}^{(-)}{}^{b}{}_{e f} 
-4 \partial^{c}\overline{\omega}^{(-)}{}^{b d e} \overline{\omega}^{(-)}{}^{a}{}_{d f} \overline{\omega}^{(-)}_{\;\;c e}{}^{f}
+ 2 \partial^{b}\overline{\omega}^{(-)}{}^{c d e} \overline{\omega}^{(-)}{}^{a}{}_{d f} \overline{\omega}^{(-)}_{\;\; c e}{}^{f}
\cr
&&
+ 2 \partial^{c}\overline{\omega}^{(-)}{}^{b e f} \overline{\omega}^{(-)}_{\; d e f} \overline{\omega}^{(+)}_{\;\;\;c}{}^{d a} 
 - \overline{\omega}^{(-)}_{\;\;\;e}{}^{c d} \overline{\omega}^{(-)}_{\; f c d} \overline{\omega}^{(+)}_{\;\;\; g}{}^{f a} \overline{\omega}^{(-)}{}^{g b e} 
 + 2 \overline{\omega}^{(-)}{}^{a c}{}^{d} \overline{\omega}^{(-)}_{\;\;g e}{}^{f} \overline{\omega}^{(-)}{}^{g}{}_{d f} \overline{\omega}^{(-)}{}^{b}{}_{c}{}^{e}\cr
&&
 - 2 \overline{\omega}^{(-)}{}^{d}{}_{c}{}^{g}
 \overline{\omega}^{(-)}_{\;\;d e}{}^{f}
 \overline{\omega}^{(-)}{}^{a}{}_{g f} \overline{\omega}^{(-)}{}^{b c e} 
 + 1/2 \overline{\omega}^{(-)}{}^{g}{}_{c d} \overline{\omega}^{(-)}{}^{a e f} \overline{\omega}^{(-)}_{\; g e f} \overline{\omega}^{(-)}{}^{b c d}\cr
 &&
- 2 \overline{\omega}^{(-)}_{\; f c}{}^{d} \overline{\omega}^{(-)}_{\;g e d} \overline{\omega}^{(-)}{}^{b c e} \left(2\overline{\omega}^{(+)}{}^{[a f g]}+\overline{\omega}^{(-)}{}^{[a f g]} \right)
\left.\vphantom{2^{\frac12}}\right]
\cr
&+& 
\frac{b^2}{8}\; \overline{e}^{(m}{}_{a} \overline{e}^{n)}{}_{b} \left[\vphantom{2^{\frac12}}\right.  \partial^{c}\overline{\omega}^{(+)}{}^{a e f} \partial_{c} \overline{\omega}^{(+)}{}^{b}{}_{e f} 
-4 \partial^{c}\overline{\omega}^{(+)}{}^{b e d} \overline{\omega}^{(+)}{}^{a}{}_{e f} \overline{\omega}^{(+)}{}_{c d}{}^{f}
+ 2 \partial^{b}\overline{\omega}^{(+)}{}^{d c e} \overline{\omega}^{(+)}{}^{a}{}_{c f} \overline{\omega}^{(+)}{}_{d e}{}^{f} \cr
&&
+ 2 \partial^{c}\overline{\omega}^{(+)}{}^{b d e} \overline{\omega}^{(+)}{}^{g}{}_{d e} \overline{\omega}^{(-)}{}_{c g}{}^{a}
 + \overline{\omega}^{(+)}{}_{e}{}^{c d} \overline{\omega}^{(+)}{}^{g}{}_{c d} \overline{\omega}^{(-)}{}_{f g}{}^{a} \overline{\omega}^{(-)}{}^{f e b}
 + 2 \overline{\omega}^{(+)}{}^{a c}{}^{f} \overline{\omega}^{(+)}{}^{g}{}_{e}{}^{d} \overline{\omega}^{(+)}{}_{g f d} \overline{\omega}^{(+)}{}^{b}{}_{c}{}^{e}\cr
&&
 - 2 \overline{\omega}^{(+)}{}^{d}{}_{c}{}^{g} \overline{\omega}^{(+)}{}_{d e}{}^{f} \overline{\omega}^{(+)}{}^{a}{}_{g f} \overline{\omega}^{(+)}{}^{b c e} 
+ 1/2 \overline{\omega}^{(+)}{}^{e}{}_{c d} \overline{\omega}^{(+)}{}^{a g f} \overline{\omega}^{(+)}{}_{e g f} \overline{\omega}^{(+)}{}^{b c d}\cr
&&-2 \overline{\omega}^{(+)}{}_{d c}{}^{f} \overline{\omega}^{(+)}{}_{g e f} \overline{\omega}^{(+)}{}^{b c e} \left(2\overline{\omega}^{(-)}{}^{[a d g]} + \overline{\omega}^{(+)}{}^{[a d g]}\right) \left.\vphantom{2^{\frac12}}\right]
\cr
&+& \frac{a b}{8}\;\overline{e}^{(m}{}_{a} \overline{e}^{n)}{}_{b} \left[\vphantom{{\frac12}^{\frac12}}\right.    
 -\partial^{c}\overline{\omega}^{(+)}{}^{d e f} \overline{\omega}^{(-)}{}^{a}{}_{c d} \overline{\omega}^{(+)}{}_{b e f}
  -\partial^{c}\overline{\omega}^{(-)}{}^{d e f} \overline{\omega}^{(+)}{}^{b}{}_{c d} \overline{\omega}^{(-)}{}^{a}{}_{e f}\cr 
 &&
 +  \partial^{b}\overline{\omega}^{(+)}{}^{c e f} \overline{\omega}^{(+)}{}^{d}{}_{e f} \overline{\omega}^{(-)}{}^{a}{}_{c d}
 +  \partial^{b}\overline{\omega}^{(-)}{}^{e c d} \overline{\omega}^{(-)}{}^{f}{}_{c d} \overline{\omega}^{(+)}{}^{a}{}_{e f}\cr
 &&
-  \frac{1}{2}  \left(2\overline{\omega}^{(-)}{}^{[d g c]}+ \overline{\omega}^{(+)}{}^{[d g c]}\right) \overline{\omega}^{(-)}{}^{a}{}_{d g} \overline{\omega}^{(+)}{}^{b e f} \overline{\omega}^{(+)}{}_{c e f}
+ \overline{\omega}^{(-)}{}^{a d c} \overline{\omega}^{(+)}{}^{b e f} \overline{\omega}^{(+)}{}_{d}{}_{e}{}^{g} \overline{\omega}^{(+)}{}_{c f g}\cr
&&
-  \frac{1}{2}  \left(2\overline{\omega}^{(+)}{}^{[e f g]}+\overline{\omega}^{(-)}{}^{[e f g]}\right) \overline{\omega}^{(+)}{}^{b}{}_{e f} \overline{\omega}^{(-)}{}^{a c d} \overline{\omega}^{(-)}{}_{g c d}
+ \overline{\omega}^{(+)}{}^{b f e} \overline{\omega}^{(-)}{}^{a c d} \overline{\omega}^{(-)}{}_{f}{}_{c}{}^{g} \overline{\omega}^{(-)}{}_{e d g} \cr
&& -  \overline{\omega}^{(+)}{}^{c e f} \overline{\omega}^{(+)}{}^{g}{}_{e f} \overline{\omega}^{(-)}{}^{b}{}_{c}{}^{d} \overline{\omega}^{(-)}{}^{a}{}_{g d} 
 -  \overline{\omega}^{(-)}{}^{g c d} \overline{\omega}^{(-)}{}^{e}{}_{c d} \overline{\omega}^{(+)}{}^{b}{}_{g}{}^{f} \overline{\omega}^{(+)}{}^{a}{}_{e f}
  \left.\vphantom{{\frac12}^{\frac12}}\right]\,.\label{gmapalph2}
\end{eqnarray}

We have explicitly verified that (\ref{gmapalph2}) is in perfect agreement with equation (4.42), (4.48) and (4.50) of \cite{Baron:2020xel}. The precise matching is straightforward but tedious, because the proposed map here sets the physical metric in terms of spin connections associated to the duality covariant frame $\overline{e}$, while in \cite{Baron:2020xel}, was obtained the inverse map. The first line in (\ref{gmapalph2}) is easily verified because $\omega^{(\pm)}$ and $\overline{\omega}^{(\pm)}$ agree at leading order, but the remaining terms require more work. We have used Cadabra software in order to verify the matching \cite{Peeters:2007wn}.

\section{Outlook}
Symmetries play a central role in the structure of effective theories in string theory. Nevertheless, not all of them can be made explicit simultaneously. That is precisely the case with T-duality and Lorentz transformations, while the former is linearly realized in DFT, the $GL(D)$ decomposition of the duality multiplets $\overline{\phi},\; \overline{g}_{mn}$ and $\overline{b}_{mn}$ transform in a very complicated way under Lorentz transformations. On the other hand, SUGRA variables $\phi,\;g_{mn}$ and $b_{mn}$ transform nonlinearly under T-duality. 

Recently different approaches, within the context of DFT, have used T-duality to get insight on possible interactions beyond quadratic order. It is therefore desirable to contrast all these results with $e.g.$ scattering amplitude or beta function computations. Having control on the map that exchange between both DFT and SUGRA schemes is a fundamental ingredient for that purpose. 

In this paper we propose a map for $g_{mn}$ and $\phi$ in terms of duality multiplets of the two-parameter deformation of DFT introduced at ${\cal O}(\alpha')$ in \cite{Marques:2015vua} and extended to all orders in \cite{Baron:2018lve}-\cite{Baron:2020xel}. This proposal was successfully contrasted with previous results up to ${\cal O}(\alpha'{}^2)$.

Unfortunately, the argument employed here cannot be extended to cover the case of the two-form. It would be interesting to shed light on this point in the near future. 

We have proved that there is a basis in SUGRA where (derivatives of the) dilaton can be completely eliminated in the Lagrangian beyond quadratic order, at least for those interactions related to the biparametric DFT. Notably, there is not a dilaton free basis in the latter. It is probably related with the fact that in order to linearly realize T-duality at higher orders, it is required to further extend the Lorentz group of SUGRA into a double Lorentz group. For instance, a way to eliminate dilaton interactions is possible if we partially fix it by identifying \footnote{In this case there is no distinction between $F_{\underline{a}}$ and $F_{\overline{a}}$ and we can also cancel terms like $F_{\underline{a}} F_{\underline{b}}(\equiv F_{\overline{a}} F_{\underline{b}})$ by using the $e.o.m.$} $\underline{e}^{m}{}_{\underline{a}}=\overline{e}^{m}{}_{\overline{b}} \delta^{\overline{b}}{}_{\underline{a}}$ , but as we said the full double Lorentz group is a crucial ingredient to linearly realized T-duality. 

Another relevant question that deserves attention is the possibility of getting iterative expressions for the equations of motion of the deformed DFT, these together with the previously mentioned maps are relevant in the context of Generalized T-dualities regarding $\alpha'$ corrections on solution generating techniques and to simplify the calculation of beta functions of sigma models (see for instance \cite{Hassler}-\cite{Hassler:2020wnp}).

The field redefinitions found in \cite{Baron:2020xel} were obtained directly from the $GGS$ transformations, whose definition is supported by the explicit verification of the closure of the algebra and by the fact that they reduce to the known $GGS$ transformations at ${\cal O}(\alpha')$. Here instead the map is derived from the action obtained in \cite{Baron:2020xel}. Even though both the $GGS$ and the action follow from the $GBdRi$,  they were obtained from two independent computations and so the agreement not only can be interpreted as a support for the map but also for the consistency of the action, if we interpret the closure of the $GGS$ as an independent proof of consistency. 
Further verifications in favor of the action found on $loc.\;cit.$ were explored in \cite{Hronek:2021nqk}, this time regarding certain pure gravity couplings.

\section*{Acknowledgments} Our work is supported by CONICET. WB receives financial support from ANPCyT-FONCyT (PICT-2015-1525) and the Grant PIP-UE B\'usqueda de nueva f\'isica.


\begin{thebibliography}{99}

\bibitem{Sen:1991zi}
A.~Sen,
``O(d) x O(d) symmetry of the space of cosmological solutions in string theory, scale factor duality and two-dimensional black holes,''
Phys. Lett. B \textbf{271} (1991), 295-300
doi:10.1016/0370-2693(91)90090-D

\bibitem{DFT}
W.~Siegel,
  ``Superspace duality in low-energy superstrings,''
  Phys.\ Rev.\ D {\bf 48} (1993) 2826
  doi:10.1103/PhysRevD.48.2826
  [hep-th/9305073].

  W.~Siegel,
  ``Two vierbein formalism for string inspired axionic gravity,''
  Phys.\ Rev.\ D {\bf 47} (1993) 5453
  doi:10.1103/PhysRevD.47.5453
  [hep-th/9302036].
  
  C.~Hull and B.~Zwiebach,
  ``Double Field Theory,''
  JHEP {\bf 0909} (2009) 099
  doi:10.1088/1126-6708/2009/09/099
  [arXiv:0904.4664 [hep-th]].

   O.~Hohm, C.~Hull and B.~Zwiebach,
  ``Background independent action for double field theory,''
  JHEP {\bf 1007} (2010) 016
  doi:10.1007/JHEP07(2010)016
  [arXiv:1003.5027 [hep-th]].
  
  \bibitem{Lescano:2021lup}
E.~Lescano,
``$\alpha'$-corrections and their double formulation,''
[arXiv:2108.12246 [hep-th]].

 \bibitem{Godazgar}
H.~Godazgar and M.~Godazgar,
  ``Duality completion of higher derivative corrections,''
  JHEP {\bf 1309} (2013) 140
  doi:10.1007/JHEP09(2013)140
  [arXiv:1306.4918 [hep-th]].
  
    \bibitem{HSZ}
  O.~Hohm, W.~Siegel and B.~Zwiebach,
  ``Doubled $\alpha'$-geometry,''
  JHEP {\bf 1402} (2014) 065
  doi:10.1007/JHEP02(2014)065
  [arXiv:1306.2970 [hep-th]].

  O.~Hohm and B.~Zwiebach,
  ``Double field theory at order $\alpha$',''
  JHEP {\bf 1411} (2014) 075
  doi:10.1007/JHEP11(2014)075
  [arXiv:1407.3803 [hep-th]].

  O.~Hohm and B.~Zwiebach,
  ``Green-Schwarz mechanism and $\alpha$'-deformed Courant brackets,''
  JHEP {\bf 1501} (2015) 012
  doi:10.1007/JHEP01(2015)012
  [arXiv:1407.0708 [hep-th]].
  
 \bibitem{Marques:2015vua}
  D.~Marques and C.~A.~Nunez,
  ``T-duality and $\alpha$'-corrections,''
  JHEP {\bf 1510} (2015) 084
  doi:10.1007/JHEP10(2015)084
  [arXiv:1507.00652 [hep-th]].
 
  \bibitem{OddStory}
  W.~H.~Baron, J.~J.~Fernandez-Melgarejo, D.~Marques and C.~Nunez,
  ``The Odd story of $\alpha$'-corrections,''
  JHEP {\bf 1704} (2017) 078
  doi:10.1007/JHEP04(2017)078
  [arXiv:1702.05489 [hep-th]].
  
 \bibitem{GSS} G.~Aldazabal, W.~Baron, D.~Marques and C.~Nunez, ``The effective action of Double Field Theory,''  JHEP {\bf 1111} (2011) 052 [JHEP {\bf 1111} (2011) 109]
  [arXiv:1109.0290 [hep-th]].

  D.~Geissbuhler, ``Double Field Theory and N=4 Gauged Supergravity,''  JHEP {\bf 1111} (2011) 116
  [arXiv:1109.4280 [hep-th]].

  \bibitem{OtherApproaches}  
  O.~A.~Bedoya, D.~Marques and C.~Nunez,
  ``Heterotic $\alpha$'-corrections in Double Field Theory,''
  JHEP {\bf 1412} (2014) 074
  doi:10.1007/JHEP12(2014)074
  [arXiv:1407.0365 [hep-th]].
  
  K.~Lee,
  ``Quadratic $\alpha$'-corrections to heterotic double field theory,''
  Nucl.\ Phys.\ B {\bf 899} (2015) 594
  doi:10.1016/j.nuclphysb.2015.08.013
  [arXiv:1504.00149 [hep-th]].
  
  \bibitem{Cosmology}
  O.~Hohm and B.~Zwiebach,
``Non-perturbative de Sitter vacua via $\alpha'$ corrections,''
Int. J. Mod. Phys. D \textbf{28} (2019) no.14, 1943002
doi:10.1142/S0218271819430028 
[arXiv:1905.06583 [hep-th]]
  
  O.~Hohm and B.~Zwiebach,
``Duality invariant cosmology to all orders in $\alpha$',''
Phys. Rev. D \textbf{100} (2019) no.12, 126011
doi:10.1103/PhysRevD.100.126011
[arXiv:1905.06963 [hep-th]].

C.~Krishnan,
``de Sitter, $\alpha'$-Corrections and Duality Invariant Cosmology,''
JCAP \textbf{10} (2019), 009
doi:10.1088/1475-7516/2019/10/009
[arXiv:1906.09257 [hep-th]].

P.~Wang, H.~Wu, H.~Yang and S.~Ying,
``Non-singular string cosmology via $\alpha^{\prime}$ corrections,''
JHEP \textbf{10} (2019), 263
doi:10.1007/JHEP10(2019)263
[arXiv:1909.00830 [hep-th]].

H.~Bernardo, R.~Brandenberger and G.~Franzmann,
``O$(d,d)$ covariant string cosmology to all orders in $\alpha^{\prime}$,''
JHEP \textbf{02} (2020), 178
doi:10.1007/JHEP02(2020)178
[arXiv:1911.00088 [hep-th]].

H.~Bernardo and G.~Franzmann,
``$\alpha'$-Cosmology: solutions and stability analysis,''
JHEP \textbf{05} (2020), 073
doi:10.1007/JHEP05(2020)073
[arXiv:2002.09856 [hep-th]].

H.~Bernardo, R.~Brandenberger and G.~Franzmann,
``String Cosmology backgrounds from Classical String Geometry,''
Phys. Rev. D \textbf{103} (2021) no.4, 043540
doi:10.1103/PhysRevD.103.043540 
[arXiv:2005.08324 [hep-th]].

H.~Bernardo, R.~Brandenberger and G.~Franzmann,
``Solution of the Size and Horizon Problems from Classical String Geometry,''
JHEP \textbf{10} (2020), 155
doi:10.1007/JHEP10(2020)155 
[arXiv:2007.14096 [hep-th]].

C.~A.~N\'u\~nez and F.~E.~Rost,
``New non-perturbative de Sitter vacua in $\alpha'$-complete cosmology,''
JHEP \textbf{03} (2021), 007
[arXiv:2011.10091 [hep-th]].

  T.~Codina, O.~Hohm and D.~Marques,
``General string cosmologies at order \ensuremath{\alpha}'3,''
Phys. Rev. D \textbf{104} (2021) no.10, 106007
doi:10.1103/PhysRevD.104.106007
[arXiv:2107.00053 [hep-th]].

  \bibitem{Baron:2018lve}
W.~H.~Baron, E.~Lescano and D.~Marqués, ``The generalized Bergshoeff-de Roo identification,''
JHEP \textbf{11} (2018), 160
doi:10.1007/JHEP11(2018)160
[arXiv:1810.01427 [hep-th]].

\bibitem{Baron:2020xel}
W.~Baron and D.~Marques,
``The generalized Bergshoeff-de Roo identification. Part II,''
JHEP \textbf{01} (2021), 171
doi:10.1007/JHEP01(2021)171
[arXiv:2009.07291 [hep-th]].

\bibitem{Green:1984sg}
M.~B.~Green and J.~H.~Schwarz,
``Anomaly Cancellation in Supersymmetric D=10 Gauge Theory and Superstring Theory,''
Phys. Lett. B \textbf{149} (1984), 117-122
doi:10.1016/0370-2693(84)91565-X

\bibitem{BdR}
  E.~Bergshoeff and M.~de Roo,
  ``Supersymmetric Chern-simons Terms in Ten-dimensions,''
  Phys.\ Lett.\ B {\bf 218} (1989) 210.
  doi:10.1016/0370-2693(89)91420-2

  E.~A.~Bergshoeff and M.~de Roo,
  ``The Quartic Effective Action of the Heterotic String and Supersymmetry,''
  Nucl.\ Phys.\ B {\bf 328} (1989) 439.
  doi:10.1016/0550-3213(89)90336-2

\bibitem{Hronek:2021nqk}
S.~Hronek and L.~Wulff,
``String theory at order \ensuremath{\alpha}'$^{2}$ and the generalized Bergshoeff-de Roo identification,''
JHEP \textbf{11} (2021), 186
doi:10.1007/JHEP11(2021)186
[arXiv:2109.12200 [hep-th]].

\bibitem{FrameDFT}
O.~Hohm and S.~K.~Kwak,
``Double Field Theory Formulation of Heterotic Strings,''
JHEP \textbf{06} (2011), 096
doi:10.1007/JHEP06(2011)096
[arXiv:1103.2136 [hep-th]].

 D.~Geissbuhler, D.~Marques, C.~Nunez and V.~Penas,
  ``Exploring Double Field Theory,''
  JHEP {\bf 1306} (2013) 101
  doi:10.1007/JHEP06(2013)101
  [arXiv:1304.1472 [hep-th]].

\bibitem{Peeters:2007wn}
K.~Peeters,
``Introducing Cadabra: A Symbolic computer algebra system for field theory problems,''
[arXiv:hep-th/0701238 [hep-th]].

\bibitem{Hassler} 
F.~Hassler and T.~Rochais,
``$\alpha'$-Corrected Poisson-Lie T-Duality,''
Fortsch. Phys. \textbf{68} (2020) no.9, 2000063
doi:10.1002/prop.202000063
[arXiv:2007.07897 [hep-th]].

\bibitem{BorsatoWulff}
R.~Borsato and L.~Wulff,
``Quantum Correction to Generalized $T$ Dualities,''
Phys. Rev. Lett. \textbf{125} (2020) no.20, 201603
doi:10.1103/PhysRevLett.125.201603
[arXiv:2007.07902 [hep-th]].

\bibitem{CodinaMarques}
T.~Codina and D.~Marques,
``Generalized Dualities and Higher Derivatives,''
JHEP \textbf{10} (2020), 002
doi:10.1007/JHEP10(2020)002
[arXiv:2007.09494 [hep-th]].

\bibitem{Hassler:2020wnp}
F.~Hassler and T.~B.~Rochais,
``O(D,D)-covariant two-loop \ensuremath{\beta}-functions and Poisson-Lie T-duality,''
JHEP \textbf{10} (2021), 210
doi:10.1007/JHEP10(2021)210
[arXiv:2011.15130 [hep-th]].
\end{thebibliography}
\end{document}